\documentstyle[aps,preprint,epsf,graphicx]{revtex}

\newcommand{\nn}{\nonumber}
\newcommand{\be}{\begin{equation}}
\newcommand{\ee}{\end{equation}}
\newcommand{\bea}{\begin{eqnarray}}
\newcommand{\eea}{\end{eqnarray}}
\newcommand{\no}{\noindent}

\newcommand{\sla}{\! \not \!}

\newcommand{\Tr}{{\rm Tr}}
\newcommand{\anu}{\bar\nu}
\newcommand{\omnu}{\omega_{\nu}}
\newcommand{\ep}{\varepsilon}
\newcommand{\bp}{\bbox{p}}
\newcommand{\bP}{\bbox{P}}
\newcommand{\bq}{\bbox{q}}
\newcommand{\bk}{\bbox{k}}

\newcommand{\bsigma}{\bbox{\sigma}}
\newcommand{\Real}{\Re{\rm e}}
\newcommand{\Img}{\Im{\rm m}}

\tighten

\begin{document}


\title{Coherent neutrino radiation in supernovae at two loops}

\author{A. Sedrakian and A. E. L. Dieperink}

\address{
        Kernfysisch Versneller Instituut,
        NL-9747 AA Groningen,
        The Netherlands
        }

\maketitle
\begin{abstract}
We develop a  neutrino transport theory, in terms of the real-time
non-equilibrium Green's functions, which
is applicable to physical conditions arbitrary far from
thermal equilibrium.
We compute the coherent neutrino radiation
in cores of supernovae by evaluating the
two-particle-two-hole (2p-2h) polarization function
with dressed propagators. The propagator dressing
is carried out in the particle-particle channel to all orders
in the  interaction.
We show that at two loops
there are two distinct sources of coherence effects in the bremsstrahlung.
One is the generically off-shell intermediate state propagation,
which leads to the Landau-Pomeranchuk-Migdal type
suppression of radiation.
We extend  previous perturbative results, obtained in the leading
order in quasiparticle width,  by deriving the
exact non-perturbative  expression.
A new contribution due to off-shell
finial/initial baryon states is treated in the
leading order in the quasiparticle width. The latter contribution
corresponds to processes of higher order than second order in the
virial expansion in the  number of quasiparticles.
At 2p-2h level, the time component of the polarization tensor
for the vector transitions vanishes identically in the soft
neutrino approximation. Vector current  thereby is conserved.
The contraction of the neutral
axial vector current with tensor interaction among the baryons
leads to a  non-vanishing contribution to the bremsstrahlung rate.
These rates  are  evaluated numerically for finite temperature pure
neutron matter at and above the nuclear saturation density.
\end{abstract}


\newpage

\section{Introduction}

Neutrino production in baryon encounters is among the
fundamental processes by which compact stars lose their
energy. The reactions  can be arranged, in general,
according to the number of participating baryons,
as the phase space  arguments play the central role in
controlling their temperature and density
dependence\cite{CHIU_SALPETER,BAHCALL_WOLF,FLOWERS_ETAL,FRIMAN_MAXWELL,VOSKRESENSKY,PETHICK}.
In the case of neutrino pair bremsstrahlung,
the leading order process in the density virial expansion
is the two-body reaction
\be\label{2BODY_BR}
B_1+B_2\to B_1+B_2+\nu_f+\overline\nu_f,
\ee
where $B$ stands for a baryon, $\nu_f$ ($\bar\nu_f$) for a neutrino
(anti-neutrino)  of flavor $f = e, \mu, \tau$.
Note that the subleading order process (i.e. the one
in the absence of the spectator)
vanishes for identical particles, as an on-shell propagating
particle cannot radiate.

The matter in neutron stars
is highly degenerate for temperatures typically below a MeV
and  the elementary excitations are
quasiparticles with well-defined energy-momentum relation.  Produced
neutrinos are typically ``soft" with energies of order of
temperature. In this limit the
intermediate quasiparticle propagator diverges as $1/\omega$ and
the amplitudes of the neutrino absorption, scattering, and
emission turn out formally divergent as $1/\omega^2$.
The infrared behaviour of the in-medium rates, however,  is
dominated by the neutrino phase space  factors, rather than
the infrared divergence of the amplitudes and
the rates of the bremsstrahlung and
its space-like analogues remain finite. At the same time, at
low temperatures, the contribution from the infrared region to
the rate of the bremsstrahlung is negligible. The combined
effect of the cancellation of the infrared divergence and the vanishing
contribution from the low frequency region
makes the quasiparticle approximation to eq. (\ref{2BODY_BR})
applicable in cold neutron stars.

During the first several tens of seconds after a supernova explosion
and core collapse the temperature of the dense nuclear matter
is of the order of several tens of MeV. The neutrino
bremsstrahlung is then suppressed,
because the formation length of
the neutrino radiation is of the same order of magnitude as
the mean free path of a baryon\cite{RAFFELT_SECKEL,JANKA,RAFFELT,HANESTAD}.
The collective effects become important on the radiation
scale (i.e. the role of the spectator in the
reaction (\ref{2BODY_BR}) is taken over by the medium)
because the baryon undergoes multiple scattering during the
radiation.
The underlying mechanism is the Landau-Pomeranchuk-Migdal (LPM) quenching
of the radiation, first  introduced in the context of QED\cite{LP}.
The central role in the theory is played by the
formation length of radiation $l_{\rm f}$. If the mean-free-path of a
baryon is much larger that the formation length $l_{\rm mfp}\gg l_{\rm f}$
then the radiation  reduces to a
sum of separate radiation events, each of which is well described by
the  Bethe-Heitler spectrum. In the opposite limit
$l_{\rm mfp}\ll l_{\rm f}$ the individual scattering
events are unresolved and the
radiation spectrum takes the Bethe-Heitler form for a  single
scattering event.
In the intermediate regime, when $l_{\rm mfp}\sim l_{\rm f}$,
the radiation amplitudes for scattering off various centers
interfere destructively and the radiation is suppressed
(Landau-Pomeranchuk-Migdal effect; for
a review see refs. \cite{RAFFELT,KNOLL_VOSKRESENSKY}).

The rates of neutrino-nucleon processes are commonly expressed
through phase space   integrals over the contraction of the
weak currents with  the polarization function of the nuclear
medium. The polarization function
(or structure function) of the supernova/neutron star matter
has been subject of many
studies\cite{RAFFELT,IWAMOTO_PETHICK,SAWYER,HOROWITZ,HAENSEL,BURROWS,REDDY_ETAL}.
The modifications  of reactions rates by the spatial correlations among (on-shell)
quasiparticles have been studied
within the Fermi-liquid theory\cite{IWAMOTO_PETHICK}, the
one-boson exchange interaction theory\cite{SAWYER}, the
relativistic random phase
approximation\cite{HOROWITZ},  the variational approach\cite{HAENSEL},
and combinations thereof\cite{BURROWS,REDDY_ETAL}.
The spatial correlations tend to suppress
reaction rates in general, although their impact on the
supernova physics is model
dependent\cite{IWAMOTO_PETHICK,SAWYER,HOROWITZ,HAENSEL,BURROWS,REDDY_ETAL}.

The common strategy of incorporating the LPM effect in the
neutrino-nucleon interaction processes is to add
a quasiparticle damping in the intermediate state propagator
by replacing $\omega$ by $\omega +i\gamma$\cite{RAFFELT_SECKEL,JANKA,RAFFELT}.
In the soft neutrino limit  the vector current coupling does not
contribute by virtue of the vector current conservation (CVC)
and the net  contribution comes from
the axial-vector transitions via baryon spin-flip.
The above modification of the intermediate state
propagator then leads to an  ansatz for
the nucleon spin structure function:
$S_{\sigma}\propto \gamma_{\sigma}/(\omega^2 + \gamma_{\sigma}^2)$
\cite{RAFFELT_SECKEL,JANKA,RAFFELT},  where $\gamma_{\sigma}$
is the nucleon spin-flip collision rate.
The ansatz generalizes the  quasiparticle picture, in a semi-phenomenological
manner,  by including the temporal correlations
among the quasiparticles in the leading order in
the quasiparticle width.  The microscopic justification of this
phenomenology emerges from the various formulations
of the finite temperature quantum filed theory, e. g.
the thermo-field dynamics\cite{BRAATEN} or the closed
diagram formalism in the Schwinger-Keldysh
technique\cite{KNOLL_VOSKRESENSKY}. A microscopic computation
is not straightforward, however. For example,
the polarization function of the medium can be computed at
one-loop, including the quasiparticle width to all orders
in $\gamma$, however {\it a priori} the current conservation
is not guaranteed at this level.  The reason,
in part, is that the  ``more complicated'' higher order
in  loop expansion
diagrams contribute  at the same order as the single
loop\cite{KNOLL_VOSKRESENSKY}.

In a previous paper we carried out a microscopic computation
of the bremsstrahlung, including the LPM effect, at
the one-loop level in a formalism based on the
quasiclassical Kadanoff-Baym transport equation\cite{SD}.
Here we extend this computation to two-loops and partially modify our
approach to include the propagator and vertex renormalization on the
same footing and  to including the tensor force explicitly.
The extension to two-loops is motivated by the following.
The long
range phenomena, driven by the weaker attractive part of the
baryon-baryon interaction, are sensitive to the resummation
in the particle-hole ($ph$) channel. On the other hand,
as well known,  one should fully resum the particle-particle ($pp$)
channel to treat the hard core of the baryon-baryon interaction.
Therefore, the $ph$ channel can be treated perturbatively by expanding
in the number of particle-hole loops, while the $pp$ channel must
be treated non-perturbatively by a full resummation of the ladder diagrams.
Thus, the separation of the long range and short range phenomena
dictates the manner in which the diagrammatic expansion is carried out.
The dressing of the single particle propagators occurs in both channels and
can be treated either explicitly, say, by considering higher order
self-energies attached to a propagator, or, alternatively, by condensing
it in the width of the propagator spectral function.
As a consequence of the separation of the scales,
the short-range correlations
can be condensed in the propagator width on the scales relevant
for the long-range phenomena. The imaginary
part of a single loop in the $ph$ channel vanishes in the time-like
region of the phase space, which is relevant for the particle
production. A  finite result emerges when one dresses the  propagators
by either extending the resummation in the $ph$ channel to two and higher
loops and/or by dressing the propagators in the $pp$ channel to all orders.
Ignoring the latter resummation, i.e. using the quasiparticle
propagators in the two-loop expansion, misses a number of short-range
collective effects, such as the LPM quenching of the radiation due to
multiple scattering. On the other hand, summing only the ladders in
the $pp$ channel does not recover  the vector current conservation
in the radiation process (in, at least,  a transparent manner).
Therefore a natural choice, motivated by the separation
of the short and long range phenomena, is to truncate the
$ph$ channel at two-loops and
to resum the $pp$ channel to all orders. The situation is
reminiscent of the  parquet resummation scheme in the first
iteration, where in both channels the driving force is the bare
baryon-baryon interaction.

Early studies of the bremsstrahlung at the quasiparticle level
modelled the strong force using
the $T$-matrix interaction\cite{FLOWERS_ETAL}, the free-space
one-boson exchange interaction\cite{FRIMAN_MAXWELL}
and their in-medium modifications
\cite{VOSKRESENSKY} supplemented with a
hard core modelled in the spirit of
the Fermi-liquid theory. The explicit use
of the tensor interaction
turned out to be crucial as there are
significant cancellations among different
diagrams, and the surviving contribution
is due to a non-trivial contraction
between the operator structures of the weak and strong interactions
(tensor force)~\cite{FRIMAN_MAXWELL}.
This motivates our ansatz for the driving force in the
particle-hole ($ph$) channel of nuclear interaction,
which includes explicitly the tensor force contribution.
We do not attempt, in the present work, to go beyond the
one-pion exchange approximation
for several reasons, one being that the non-perturbative
treatment of the interaction does not change the spin, isospin,
and tensor operator structure of the interaction,
and important cancellations
in the radiation matrix elements will be preserved
in a more advanced treatment.
We also want to be able to isolate the finite width effects
in our comparisons to the earlier work done in the one-pion
exchange approximation\cite{FRIMAN_MAXWELL,RAFFELT}.
The situation is different in the particle-particle ($pp$)
channel, where the short-range correlations have to be treated in
a non-perturbative manner by summing up the ladder diagrams to
all orders. We do this in the finite-temperature Brueckner
theory.

The paper is organized as follows. In Section 2, starting from
the Kadanoff-Baym formalism, we derive a single-time
transport equation for (anti)-neutrinos with
collision integrals driven by (anti)-neutrino coupling to
baryons via the polarization tensor of the medium. The polarization
tensor is computed in the 2p-2h approximation in Section~3.
The summation of the ladder diagrams in the $pp$ channel
within the finite temperature Brueckner theory is
described in Section 4.
Section~5 evaluates the phase space integrals and
 neutrino bremsstrahlung emissivities.
The numerical results are presented in Section~6.
Section 7 summarizes our main results.

\section{Neutrino Transport Formalism}

\subsection{Neutrino propagators}

The theory of neutrino radiation can be conveniently formulated in terms
of the real-time quantum neutrino transport.
Let us start by defining the various time-ordered
Greens functions of massless Dirac neutrinos.
These can be written in the generic matrix form
  \be\label{MATRIX_GF}
   i \underline{S}_{12} =i\left( \begin{array}{cc}
                              S^{c}_{12} & S^{<}_{12} \\
                              S^{>}_{12} & S^{a}_{12}
                           \end{array} \right) =
                    \left( \begin{array}{cc}
\left <T\psi(x_1) \bar \psi(x_2)\right > & -\left <\bar\psi(x_2)\psi(x_1)\right > \\
\left <\psi(x_1)\bar\psi(x_2)\right > & \left <\tilde T\psi(x_1)\bar\psi(x_2)\right >
                               \end{array} \right)= i\left( \begin{array}{cc}
                              S^{--}_{12} & S^{-+}_{12} \\
                              S^{+-}_{12} & S^{++}_{12}
                           \end{array} \right),
  \ee
where $\psi(x)$ are the neutrino field operators, $\bar\psi = \gamma^0\psi^*$,
$T$ is the chronological time ordering operator, and $\tilde T$
is the anti-chronological time ordering operator; the indexes
$1 = x_1$, $2=x_2$,...
collectively denote the space-time and discrete quantum numbers.
The neutrino matrix propagator is further assumed to obey
the Dyson equation,
\bea\label{DYSON1}
\underline{S}(x_1,x_2) & = &\underline{S}_0(x_1,x_2)
       + \underline{S}_0(x_1,x_3)
         \underline{\Omega}(x_3,x_2) \underline{S}(x_2,x_1) \nonumber \\
         &=&\underline{S}_0(x_1,x_2) + \underline{S}(x_1,x_3)
         \underline{\Omega}(x_3,x_2) \underline{S}_0(x_2,x_1),
\eea
where $S_0(x_1,x_2)$ is the free neutrino
propagator and  $S_0^{-1}(x_1,x_2) S_0(x_1,x_2) = \sigma_z \delta(x_1-x_2)$,
$\sigma_z$ is the third component of the Pauli matrix,
$\underline\Omega$  is the neutrino proper self-energy and
we assume  integration (summation) over the repeated variables.
The self-energy $\underline\Omega$ is a $2\times 2$ matrix with elements
defined on the contour in terms of the Dyson equation.
The quasiclassical neutrino transport equation follows from
the Dyson equation in the `conjugate subtracted' form\cite{KADANOFF_BAYM,MALFLIET}:
  \bea
   i\underline{S}(x_1,x_2)  \not\!\partial_{x_2}
    -i \not\!\partial_{x_1}
       \underline{S}(x_1,x_2) =
                      \underline{S}(x_1,x_3)
                               \underline{\Omega}(x_3,x_2)
                               \underline{ \sigma_{z} }
                              -\underline{ \sigma_{z} }
                               \underline{\Omega}(x_1,x_3)
                                \underline{S}(x_3,x_2) ,
 \label{DYSON}
 \eea
Note that the initial correlations are neglected in eq. (\ref{DYSON}).
The set of the four Green's functions above can
be supplemented by the retarded and advanced Green's functions
which are defined as
\bea
i S^R_{12}=\theta(t_1-t_2)
     \langle \left\{ \psi(x_1),
       \overline{\psi}(x_2) \right\}\rangle ,\quad
   i S^A_{12}=-\theta(t_2-t_1)
     \langle \left\{ \psi(x_1),
       \overline{\psi}(x_2) \right\} \rangle ,
\eea
where $\theta(x)$ is the Heaviside step function on the
real-time contour defined as $d \theta(x)/dx = \sigma_z \delta(x)$.
The retarded and advanced Green's functions obey integral
equations in the quasiclassical limit.
The relations between the six Green's functions are
listed in the Appendix A. The transport equation
for the off-diagonal elements of the matrix
Green's function reads
 \bea
&&  \left[ \not\!\partial_{x_3} -\Real\, \Omega^R(x_1,x_3),S^{>,<}(x_3,x_2)\right]
  -\left[\Real\, S^R(x_1,x_3),\Omega^{>,<}(x_3,x_2)\right]\nonumber \\
 &&\hspace{3cm}  = \frac{1}{2}\left\{S^{>,<}(x_1,x_3),\Omega^{>,<}(x_3,x_2)\right\}
   +\frac{1}{2}\left\{\Omega^{>,<}(x_1,x_3),S^{>,<}(x_3,x_2)\right\},
\label{DYSON_OFF}
\eea
where  $[\, ,\,]$ and $\{\, ,\, \}$ stand for commutator and anti-commutator,
respectively.
In arriving at eq. (\ref{DYSON_OFF}) we assumed the existence of the Lehmann
representation for the neutrino propagators; as a results we have
$\Real~ S^R = \Real~ S^A\equiv \Real~ S$ and $\Real~ \Omega^R = \Real~ \Omega^A\equiv \Real~ \Omega$.

For the present purposes the neutrino dynamics can
be treated semiclassically, by separating the slowly varying
center-of-mass coordinates from the rapidly varying
relative coordinates. Carrying out a Fourier transform with
respect to the relative coordinates and keeping the first-order
gradients in the slow variable we arrive at a quasiclassical neutrino
transport equation
\bea
 &&  i\left\{\Real S^{-1}(q,x),S^{>,<}(q,x)\right\}_{P.B.}
  +i\left\{\Real\, S(q,x),\Omega^{>,<}(q,x)\right\}_{P.B.}  \nonumber\\
  &&\hspace{3cm}  =  S^{>,<}(q,x)\Omega^{>,<}(q,x)
   +\Omega^{>,<}(q,x)S^{>,<}(q,x),
\label{TRANS_EQ}
\eea
where $q\equiv (\bq , q_0)$
and  $x$ are the neutrino four momentum and the center-of-mass
space-time coordinate, respectively,
$\{\dots\}_{P.B.}$ is the four-dimensional Poisson bracket.
The l.h.s. of eq. (\ref{TRANS_EQ}) is the precursor of the drift term
of the Boltzmann equation. The second Poisson bracket, however, does
not fit in the Boltzmann description and can be eliminated by an expansion
of the neutrino propagator in the leading (quasi-particle) and next-to-leading
order terms in the small neutrino damping: 
$S^{>,<}(q,x)=S_0^{>,<}(q,x)+S_{1}^{>,<}(q,x)$.
A direct evaluation of the Poisson brackets decouples
 the l.h.s. of transport
equation (\ref{TRANS_EQ}) to the leading order with respect to the
small damping of neutrino/anti-neutrino states ($\Img\Omega(q,x)
/\Real\Omega(q,x) \ll 1 $). The quasiparticle part of
the transport equation
\bea\label{QPA_TRANS}
  i\left\{\Real S^{-1}(q,x),S_0^{>,<}(q,x)\right\}_{P.B.}
   =S^{>,<}(q,x)\Omega^{>,<}(q,x)+\Omega^{>,<}(q,x)S^{>,<}(q,x)
\eea
describes the evolution of the distribution
function (Wigner function) of  on-shell
excitations with the l.h.s. corresponding to the
drift term of the Boltzmann equation. The r.h.s. corresponds to
the collision integral with the self-energies $\Omega^{>,<}(q,x)$
having the meaning of the collision rates. The advantage of this form
of the (generalized) collision integral is that it admits systematic
approximations in terms of the Feynman perturbation theory. The remainder
part of the transport equation
\bea
 i\left\{\Real S^{-1}(q,x),S_{1}^{>,<}(q,x)\right\}_{P.B.}
  +i\left\{\Real\, S(q,x),\Omega^{>,<}(q,x)\right\}_{P.B.}  =  0,
\eea
relates the finite width part of the
neutrino propagator to the self-energies in a form of a local
functional which depends on the local (anti-)neutrino particle distribution
function and their coupling to the matter.

\subsection{On-shell neutrino approximation}

The on-mass-shell neutrino propagator is related to the single-time
distribution functions (Wigner functions) of neutrinos and anti-neutrinos,
$f_{\nu}(q,x)$ and $f_{\bar\nu}(q,x)$, via the ansatz
\bea
S_0^<(q,x)
&=& \frac{i\pi\sla q}{\omnu(\bq)}
    \Big[ \delta\left(q_0-\omnu(\bq)\right)f_{\nu}(q, x)
-\delta\left(q_0+\omnu(\bq)\right) \left(1-f_{\bar \nu}
(-q,x)\right) \Big],
\eea
where
$\omnu(\bq)=c\vert q\vert$ is the on-mass-shell
neutrino/anti-neutrino energy.  Note that the ansatz
includes {\it simultaneously} the neutrino particle states and
anti-neutrino hole states, which propagate in, say, positive time
direction. Similarly, the on-shell propagator
\bea
S_0^>(q,x)
&=& - \frac{i\pi\sla q}{\omnu(\bq)}
    \Big[ \delta\left(q_0-\omnu(\bq)\right)
    \left(1-f_{\nu}(q,x)\right)
-\delta\left(q_0+\omnu(\bq)\right)f_{\bar\nu} (-q,x)\Big],
\eea
corresponds to the states propagating in the reversed time
direction and, hence,
includes the anti-neutrino particle states and
neutrino hole states.

To recover the Boltzmann drift term in the on-shell limit,
we take the trace on both
sides of the transport equation  (\ref{TRANS_EQ})
and integrate over the (anti-)neutrino energy $q_0$.
The first term on l.h.s. of  eq. (\ref{TRANS_EQ}) reduces then to the
drift term of the Boltzmann equation.
The single time Boltzmann equation (hereafter BE)
for neutrinos is obtained after integrating over the
positive energy range:
\bea\label{BE_NU}
& & \left[\partial_t + \vec \partial_q\,\omnu (\bq) \vec\partial_x
\right] f_{\nu}(\bq,x) =
\int_{0}^\infty \frac{dq_0}{2\pi} {\rm Tr} \left[\Omega^<(q,x)S_0^>(q,x)
-\Omega^>(q,x)S_0^<(q,x)\right];
\eea
a similar equation follows for the  anti-neutrinos if one integrates
in eq. (\ref{TRANS_EQ})
over the range $[-\infty , 0]$.

\begin{center}
\includegraphics[height=.8in,width=3.2in]{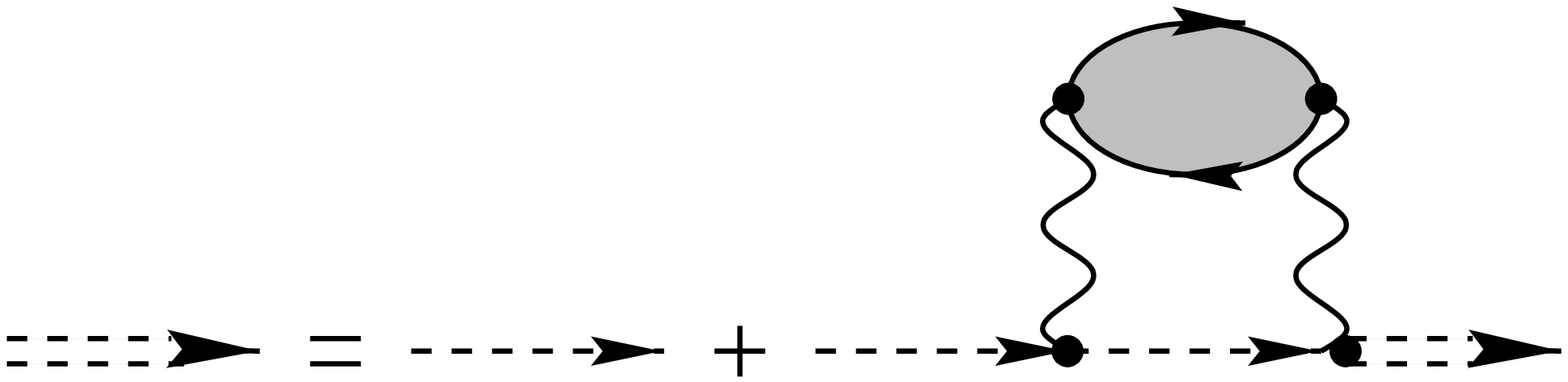}
\end{center}
{\footnotesize{Fig 1: The neutrino Dyson equation in terms of
the Feynman diagrams. The dashed curve corresponds to the $S$-propagator,
which includes the neutrinos and anti-neutrino holes moving in the
same time direction; (reverting the time-direction one finds the
Dyson equation for anti-neutrinos and neutrino holes).
The shaded loop is the baryon
polarization tensor. The wavy lines correspond to the $W^{\pm},Z^0$ boson
propagators.}}
\label{fig1}

\no
The different energy integration limits select from the r.h.s. of
the transport equations the  processes leading to
modifications of the distribution functions of (anti-)neutrinos.
The separation of the transport equation into neutrino and anti-neutrino
parts is arbitrary, however is motivated by the observation
that the  fundamental quantities of neutrino radiative transport, as the
energy densities or neutrino fluxes, can be obtained by taking the
appropriate moments of BEs. These quantities are not symmetric with
respect to the neutrino/anti-neutrino populations in general. E.g.
the neutrino emissivities (energy output per unit time per unit volume)
for processes based on $\beta$-decay reactions are given by the
zeroth order moment of the anti-neutrino BE, and it is sufficient
to consider only the BE for anti-neutrinos.
In the case of the bremsstrahlung
we have to eventually sum  these equations; still the relation of the
transport self-energies to particular processes becomes transparent
if one treats the transport equations separately.

\subsection{Collision integrals}

We adopt the standard model for the description of
the neutrino-baryon interactions
and write the neutral current interaction  Hamiltonian in the from:
\bea\label{HAM}
H_{\rm int} = \frac{G}{2\sqrt{2}} \Gamma^H \, \Gamma^L, \quad
\Gamma^H =\overline \phi \gamma_{\mu}
(c_V - c_A\gamma_5) \phi ,\quad \Gamma^L
= \overline\psi \gamma^{\mu}(1-\gamma_5)
\psi,
\eea
where  $G$ is the weak coupling constant,  $\psi$ and $\phi$ are the
neutrino and baryon field operators, $c_V$ and $c_A$ are the
dimensionless weak neutral-current vector and axial vector
coupling constants.

The diagrams contributing to the neutrino emission rates
can be arranged in a perturbation expansion with respect
to the weak interaction. The lowest order in the weak interaction
Feynman diagrams which contribute to scattering,
emission, and absorption processes are shown in the Fig. 1.
The corresponding transport self-energies are read-off from
the diagram
\bea
-i\Omega^{>,<}(q_1,x) &=& \int \frac{d^4 q}{(2\pi)^4}
\frac{d^4 q_2}{(2\pi)^4}(2\pi)^4 \delta^4(q_1 - q_2 - q)
i\Gamma_{L\, q}^{\mu}\, iS_0^{<}(q_2,x) i\Gamma_{L\, q}^{\dagger\, \lambda}
i \Pi^{>,<}_{\mu\lambda}(q,x),
\eea
where $\Pi^{>,<}_{\mu\lambda}(q)$ are the off-diagonal elements
of the matrix of the baryon polarization tensor,
$\Gamma_{L\, q}^{\mu}$ is the weak interaction vertex.
The contact interaction (\ref{HAM}) can be used
for the energy-momentum
transfers much smaller than the vector boson mass, $q\ll m_Z, m_W$.
Let us first concentrate on the BE for neutrinos. Define the loss and
gain terms of the collision integral as:
\bea
I_{\nu}^{>,<}(\bq,x)=\int_{0}^\infty \frac{dq_0}{2\pi} {\rm Tr}
\left[\Omega^{>,<}(q,x)S_0^{>,<}(q,x)\right].
\eea
Substituting the self-energies and the propagators in the collision
integrals we find for, e.g., the gain part:
\bea\label{GAIN}
I_{\nu}^{\rm <}(\bq_1,x)&=& -i\int_{0}^\infty \frac{dq_{10}}{2\pi}
 {\rm Tr}
\Biggl\{ \int_{-\infty}^{\infty} \frac{d^4 q}{(2\pi)^4}
\frac{d^4 q_2}{(2\pi)^4}(2\pi)^4\delta^4(q_1 - q_2 - q)
\Gamma^{\mu}_L\frac{\pi\sla q_2}{\omnu(\bq_2)}
\Big[\delta\left(q_{02}-\omnu( \bq_2)\right)f_{\nu}(q_2, x)\nonumber\\
&-&\delta\left(q_{02}+\omnu(\bq_2)\right) \left(1-f_{\bar \nu}
(-q_2,x)\right) \Big]\Gamma^{\dagger\,\lambda}_L
\frac{\pi\sla q_1}{\omnu(\bq_1)}
\delta\left(q_{10}-\omnu(\bq_1)\right)\left(1-f_{\nu}(q_1,x)\right)
\Pi_{\mu\lambda}^{>}(q,x)\Biggr\}. \nonumber \\
\eea
The loss term is obtained by replacing in eq. (\ref{GAIN}) the
neutrino Wigner functions by the neutrino-hole functions
$f_{\nu}(q, x) \to (1-f_{\nu}(q, x))$ and the anti-neutrino-hole
Wigner functions by the anti-neutrino  functions
$ \left(1-f_{\bar \nu}(-q,x)\right) \to f_{\bar \nu}(q,x)$.
The terms proportional $(1-f_{\nu}) f_{\nu}$ and
$(1- f_{\nu})(1-f_{\anu}) $ in the gain part of
the collision integral, $I_{\nu}^<(\bq)$, correspond to the
neutrino scattering-in and emission  contributions, respectively.
The terms  proportional  $f_{\nu} (1-f_{\nu})$ and $f_{\nu}f_{\anu}$
in the loss part of the collision integral, $I_{\nu}^>(\bq)$, are the
neutrino scattering-out and absorption contributions.

The loss and gain collision integrals for the anti-neutrinos can be defined
in a manner, similar to the case of neutrinos, with the energy integration
spanning the negative energy range
\bea
I_{\anu}^{>,<}(\bq,x)=\int^{0}_{-\infty} \frac{dq_0}{2\pi} {\rm Tr}
\left[\Omega^{>,<}(q,x)S_0^{>,<}(q,x)\right].
\eea
Using the above expressions for the
self-energy and the propagators, we find, e.g.,
for the gain term:
\bea\label{GAIN2}
I_{\anu}^{<}(\bq_1,x)&=& i\int^{0}_{-\infty}
\frac{dq_{10}}{2\pi} {\rm Tr}\Biggl\{ \int_{-\infty}^{\infty}
\frac{d^4 q}{(2\pi)^4}\frac{d^4 q_2}{(2\pi)^4}(2\pi)^4\delta^4(q_1- q_2- q)
\Gamma^{\mu}_L\frac{\pi\sla q_2}{\omnu(\bq_2)}\Big[\delta\left(q_{02}
-\omnu(\bq_2)\right)f_{\nu}(q_2, x)
\nonumber \\
&-&\delta\left(q_{02}+\omnu(\bq_2)\right)\left(1-f_{\bar \nu}(-q_2,x)\right)
\Big]\,\Gamma^{\dagger\lambda}_L\frac{\pi\sla q_1}{\omnu(\bq_1)}
\delta\left(q_{10}+\omnu(\bq_1)\right)f_{\anu}(-q_1,x)
\Pi^{>}_{\mu\lambda}(q,x)\Biggr\}.
\eea
The loss term is obtained by making replacements in eq. (\ref{GAIN2})
analogous to those applied to eq. (\ref{GAIN}).
The terms proportional $ f_{\nu}f_{\anu}$ and
$f_{\anu}(1-f_{\anu}) $ in the gain part of
the collision integral, $I_{\anu}^<(\bq)$, then correspond to the
neutrino absorption and scattering-out  contributions.
The terms  proportional $(1-f_{\anu}) (1-f_{\nu})$ and
$(1-f_{\anu})f_{\anu}$ in the loss part of the collision integral,
$I_{\anu}^>(\bq)$, are the neutrino emission and scattering-in contributions,
respectively. Note that, when the neutrinos are in a thermal
equilibrium with the baryons, the collision integrals for the
scattering-in/scattering-out and for the absorption/emission cancel.
Under the conditions of detailed balance the (anti-)neutrino distribution
function reduces to the Fermi-Dirac form.

\subsection{Bremsstrahlung emissivity}

The neutrino-pair emissivity (the power of the
energy radiated per volume unit)
is obtained by multiplying the left-hand-sides of the
neutrino and anti-neutrino
by their energies, respectively, summing the BEs, and integrating over a
phase space element:
\bea
\epsilon_{\nu\bar\nu}&=&\frac{d}{dt}\int\!\frac{d^3q}{(2\pi)^3}
\left[f_{\nu}(\bq) +f_{\bar\nu}(\bq)\right]\omnu(\bq)=
\int\!\frac{d^3q}{(2\pi)^3}\left[I_{\nu}^{<, {\rm em}}(\bq)
-I_{\anu}^{>, {\rm em}}(\bq)\right]\omnu(\bq),
\eea
where in the collision integrals we kept only the terms
which correspond to the processes with the
neutrino and anti-neutrino in the final state (bremsstrahlung)
\bea
&&\int\!\frac{d^3q_1}{(2\pi)^3}I_{\nu}^{>,<, {\rm em}}(\bq_1)\omnu(\bq_1)
= i \int\!\frac{d^3q_1}{(2\pi)^32 \omnu(\bq_1)}
\frac{d^3 q_2}{(2\pi)^3 2\omnu(\bq_2)}
\frac{d^4 q}{(2\pi)^4}(2\pi)^4 \delta^3(\bq_1+\bq_2-\bq) \nonumber\\
\label{COLL_INT1}
&&\hspace{1cm}
\delta(\omnu(\bq_1)+\omnu(\bq_2)-q_{0})\omnu(\bq_1)
\left[1-f_{\nu}(\omnu(\bq_1))\right]
\left[1-f_{\anu}(\omnu(\bq_2))\right]\Lambda^{\mu\lambda}(q_1,q_2)
\Pi_{\mu\lambda}^{>,<}(q,x),
\label{COLL_INT2}
\eea
and $\Lambda^{\mu\lambda} = {\rm Tr}\left[\gamma^{\mu}
(1 - \gamma^5)\sla q_1\gamma^{\nu}(1-\gamma^5)\sla q_2\right]$.
The collision integrals for neutrinos and anti-neutrinos can be
combined if  one uses the identities  $\Pi_{\mu\lambda}^{<}(q)
=\Pi_{\lambda\mu}^{>}(-q) = 2i g_B(q_0) {\Img}\,
\Pi_{\mu\lambda}^R(q)$; here $g_B(q_0)$ is the Bose distribution
function and $\Pi^R_{\mu\lambda}(q)$ is the retarded component
of the polarization tensor. With these modifications
the neutrino-pair bremsstrahlung emissivity becomes
\bea\label{EMISSIVITY}
\epsilon_{\nu\anu}&=& - 2\left( \frac{G}{2\sqrt{2}}\right)^2
\sum_f\int\!\frac{d^3q_2}{(2\pi)^32 \omnu(\bq_2)}
\int\!\frac{d^3 q_1}{(2\pi)^3 2\omnu(\bq_1)}
\int\!
\frac{d^4 q}{(2\pi)^4}
\nonumber\\
&&\hspace{1cm}
(2\pi)^4 \delta^3(\bq_1 + \bq_2 -  \bq)
\delta(\omnu(\bq_1)+\omnu(\bq_2)-q_{0})\, \left[\omnu(\bq_1)+\omnu(\bq_2)\right]
\nonumber\\
&&\hspace{2cm}
 g_B(q_0)\left[1-f_{\nu}(\omnu(\bq_1))\right]
\left[1-f_{\anu}(\omnu(\bq_2))\right]
 \Lambda^{\mu\lambda}(q_1,q_2){\Img}\,\Pi_{\mu\lambda}^R(q).
\eea
We note that  eq. (\ref{EMISSIVITY}) is applicable for arbitrary
deviation from equilibrium, as the equilibrium properties of the
neutrinos and baryons have not been used in the derivation (e.g.
the temperature of the bath drops out if one assumes an
initially uncorrelated state). Therefore eq.
(\ref{EMISSIVITY}) is applicable beyond the boundaries
of the linear response theory or the $S$-matrix theory which explicitly
resort to the equilibrium properties of the system as a reference
point.

\section{Two-loop Baryon polarization function}

In this section we start the implementation
of the perturbative scheme motivated in the
introduction. Our strategy is the separation
of the long and short range phenomena in the
$ph$ and $pp$ channels. Here we carry out the first step
by expanding the particle-hole channel and truncating
it at two loops. This fixes the amount of the long-range
correlations in the theory. The short-range effects are
condensed in the width of the particle-hole propagators,
which is specified in a later section by summing the ladder diagrams.

\subsection{Baryon propagators}

Although we shall treat the baryon sector in the equilibrium limit,
it is still useful to define the six Green's functions of the
non-equilibrium theory, as in the case of neutrinos.
The matrix Green's function of non-relativistic baryons
 is defined in the standard way
\be\label{MATRIX_GFB}
    i\underline{G}_{12} =i \left( \begin{array}{cc}
                              G^{c}_{12} & G^{<}_{12} \\
                              G^{>}_{12} & G^{a}_{12}
                           \end{array} \right) =
                    \left( \begin{array}{cc}
\left <T\phi(x_1)\phi^{\dagger}(x_2)\right > & -\left <\phi^{\dagger}(x_2)\phi(x_1)\right > \\
\left <\phi(x_1)\phi^{\dagger}(x_2)\right > &
\left <\tilde T\phi(x_1)\phi^{\dagger}(x_2)\right >
                               \end{array} \right)=i \left( \begin{array}{cc}
                              G^{--}_{12} & G^{-+}_{12} \\
                              G^{+-}_{12} & G^{++}_{12}
                           \end{array} \right),
\ee
where $\phi(x)$ are the baryon field operators. In terms of
these operators  the retarded and advanced function are defined as
\bea
i G^R_{12}=\theta(t_1-t_2)
     \langle \left\{ \phi(x_1),
       \phi^{\dagger} (x_2) \right\}\rangle ,\quad
i G^A_{12}=-\theta(t_2-t_1)
     \langle \left\{ \phi(x_1),
       {\phi}^{\dagger} (x_2) \right\} \rangle .
\eea
The structure of the proper self-energy matrix
$\underline{\Sigma}$ is identical to eq.
(\ref{MATRIX_GFB}) and its elements
are defined via the Dyson equation for baryons:
\bea
\underline{G}(x_1,x_2) & = &\underline{G}_{0}(x_1,x_2)
                     +\underline{G}_{0}(x_1,x_3)
         \underline{\Sigma}(x_3,x_2) \underline{G}(x_2,x_1) \nonumber \\
         &=&\underline{G}_{0}(x_1,x_2) + \underline{G}(x_1,x_3)
         \underline{\Sigma}(x_3,x_2) \underline{G}_{0}(x_2,x_1).
\eea
In a complete analogy to the neutrino sector, we approximate
the Green's functions by their quasiclassical counterparts
by defining center-of-mass and relative space-time coordinates
and  Fourier transform with respect to the
relative space-time coordinates. In the equilibrium limit
the dependence of the quasiclassical Green's functions
on their center-of-mass space-time coordinate
is trivial and can be dropped.
The distribution function of the baryons is related to the
off-diagonal elements of the matrix Green function by
the exact relations
\bea\label{ANSATZ}
-iG^<(p) = a(p) f_N(p),  \quad iG^>(p) = a(p)[1- f_N(p)],
\eea
where $a(p)=i[ G^R(p)-G^A(p)] = i[ G^>(p)-G^<(p)] $
is the baryon spectral function,
$f_N(p)=[{\rm exp}(\beta(\omega-\mu))+1]^{-1}$
is the Fermi-Dirac distribution function, $\beta = T^{-1}$ is the
inverse temperature and $\mu$ is the chemical potential
(relations (\ref{ANSATZ}) will be refereed
to as the Kadanoff-Baym ansatz in the following). The
quasiparticle energy, $\varepsilon_p = p^2/2m
+\Real\Sigma^R(p)\vert_{\omega=\ep_p}$ follows from the
solution of the Dyson equation $G^R(p)=\left[\omega-\varepsilon_p
+i\Img \Sigma^R(p) \right]^{-1}$.
When damping of quasiparticle states
is small,
$\Img \Sigma^R(p) \ll \Real \Sigma^R(p)$, the propagators can be
decomposed into quasiparticle and background contributions, e.g.,
\bea
\label{exp}
G^<(p) &\simeq& 2\pi i z(\bp)f_N(\bp)
\delta(\omega -\ep_p) -\Sigma^<(p)
\, \frac{{\cal P}}{(\omega -\ep_p)^2}+{\cal O}(\gamma^2).
\eea
Note that the self-energy appearing in the
denominator of the second term of eq.~(\ref{exp})
via the dispersion relation is restricted, to the
leading order in damping, to the mass-shell.
In equilibrium,
\bea
i\Sigma^<(p) = \gamma(p)f_N(p), \quad -i\Sigma^>(p) = \gamma(p)[1-f_N(p)],
\eea
where $\gamma(p) = -2\Img \Sigma(p)$ is the width of the 
baryon spectral function.
The wave-function renormalization, $z(\bp)$, in the same approximation is
\bea
\label{Z} z(\bp)
=1-\int\!\frac{d\omega'}{2\pi}
\Img \Sigma(\omega',\bp)\,
\frac{\cal P}{(\omega '-\omega)^2}\Big|_{\omega=\ep_p},
\eea
where we used the integro-differential form of the Kramers-Kronig relation:
\bea
\frac{d}{d \omega}{\rm \Real}\, \Sigma(\omega,\bp)&=&
\int\!\frac{d\omega'}{\pi}  {\rm  \Img}\, \Sigma(\omega',\bp)
\frac{{\cal P}}{(\omega-\omega')^2}.
\eea
On inserting the expression of the wave-function renormalization
(\ref{Z}) in the expansion (\ref{exp}) we find the final form
of the propagator
\be\label{exp1}
G^<(p) \simeq 2\pi i f_N(\bp) -2\pi i \int \frac{d\omega'}{2\pi}
\, \gamma(p')\frac{{\cal P}}{(\omega' -\ep_p)^2}
\left[\delta(\omega-\ep_p)-\delta(\omega-\omega')\right]
f_N(\omega).
\ee
Note that this form of propagator renders the strict
fulfillment of the spectral sum rule,
\be \label{sum_rule}
\int\!\frac{d\omega}{2\pi}~a(p) = 1,
\ee
at any order in the expansion with respect to the damping.

Using the linear relations among the propagators, listed
in Appendix A, we find for the causal propagator:
\bea
G^{--}(p)
&=& \frac{\omega-(\epsilon_p+\Real \Sigma(p)-\mu)}
{\left[\omega-(\epsilon_p+\Real \Sigma(p)-\mu)\right]^2+
\left[\Img\Sigma(p)\right]^2}\nonumber\\
&-&\frac{i\Img\Sigma(p)}
{\left[\omega-(\epsilon_p+\Real \Sigma(p)-\mu)\right]^2
+ \left[\Img\Sigma(p)\right]^2}
{\rm tanh}\left(\frac{\beta\omega}{2}\right),
\eea
where ${\rm tanh}\left({\omega}/{2}\right)\equiv [1-2f_N(\omega)]$
and $\epsilon_p = p^2/2m$.
As the evaluation of the baryon polarization function requires
the causal and acausal Green's functions  of the type $G^{--}(q+p)$,
we note here that, the denominator of such a function
can be expanded in the limit $vq\ll\omega$, where $v\ll 1$ is the
characteristic velocity of a baryon,
\be\label{DENOM_EXP}
(\omega +\varepsilon_p)-\varepsilon_{\vec p+\vec q}
\simeq
\omega-\bp\cdot\bq/m
-q\frac{\partial}{\partial p}~ \Real\Sigma(p)
-\epsilon_{q} \simeq\omega,
\ee
to the leading order. The approximation (\ref{DENOM_EXP}) will
be referred in the following as the {\it soft-neutrino 
approximation}.
We also employed the non-relativistic limit for baryons.
If we use the ansatz $\gamma(-\omega)=\gamma(\omega)$,
which is exact in the phenomenological Fermi-liquid theory and
will be verified in our microscopic calculations, then
\bea
G^{--}(\pm\omega,\bp) &=& \pm\frac{\omega}{\omega^2+\gamma(\omega,\bp)^2/4}
\mp i\frac{\gamma(\omega,\bp)/2}{\omega^2+\gamma(\omega,\bp)^2/4}
~{\rm tanh}\left(\frac{\beta\omega}{2}\right),      \\
-G^{++}(\pm\omega,\bp) &=& \pm\frac{\omega}{\omega^2+\gamma(\omega,\bp)^2/4}
\pm i\frac{\gamma(\omega,\bp)/2}{\omega^2+\gamma(\omega,\bp)^2/4}
~{\rm tanh}\left(\frac{\beta\omega}{2}\right),
\eea
where the second equation follows from the relation
$[G^{--}(p)]^*=-G^{++}(p)$, valid in the momentum representation (see
Appendix A).
Thus  both propagators are odd under the exchange of the sign
of $\omega$, a property which will be important in establishing
the vector current conservation in the radiation processes
discussed below. Since the dependence of the
the quasiparticle width on the momentum is weak in the density and
temperature range of interest it is useful
to define momentum average quasiparticle width which a function
only of the frequency. This approximation is implemented in the
phase space integrations below.

\subsection{The interactions}

The central ingredient of a bremsstrahlung
process is the modelling of strong the interaction. For the
particle-hole interaction a reasonable, but
not unique, choice is the one-pion exchange interaction combined
with  a contact  interaction in the spirit of the Fermi-liquid
theory:
\be\label{VBB}
V_{[ph]}(k) = \left(\frac{f_{\pi}}{m_{\pi}}\right)^2
\left(\bsigma_1\cdot \bk \right)D^{--}(\bk)\left(\bsigma_2\cdot \bk \right)
+ f_0 + f_1 (\bsigma\cdot\bsigma),
\ee
where $f_{\pi}$ is the pion decay constant, $m_{\pi}$ is the pion mass,
$D^{--}(\bk)$ is the one-shell causal pion propagator,
$f_0$ and $f_1$ are the coupling parameters of the Fermi-liquid
theory, $\bsigma$  is the vector of the Pauli matrices.
The non-relativistic reduction of the neutrino-neutron 
interaction vertex (\ref{HAM}) is
\be
\Gamma_{\mu}^H=-\frac{G}{2\sqrt{2}}
\left(\delta_{\mu 0} - g_A\delta_{\mu i}\sigma_i\right),
\ee
where $g_A=1.25$ is the axial-vector coupling constant.

\subsection{Direct contribution to the polarization function}

The three topologically different {\it direct} diagrams (i.e.
those which do not involve an exchange of outgoing
particles) are shown in Fig. 2a-c.

\begin{center}
\includegraphics[height=1.in,width=5in,angle=0]{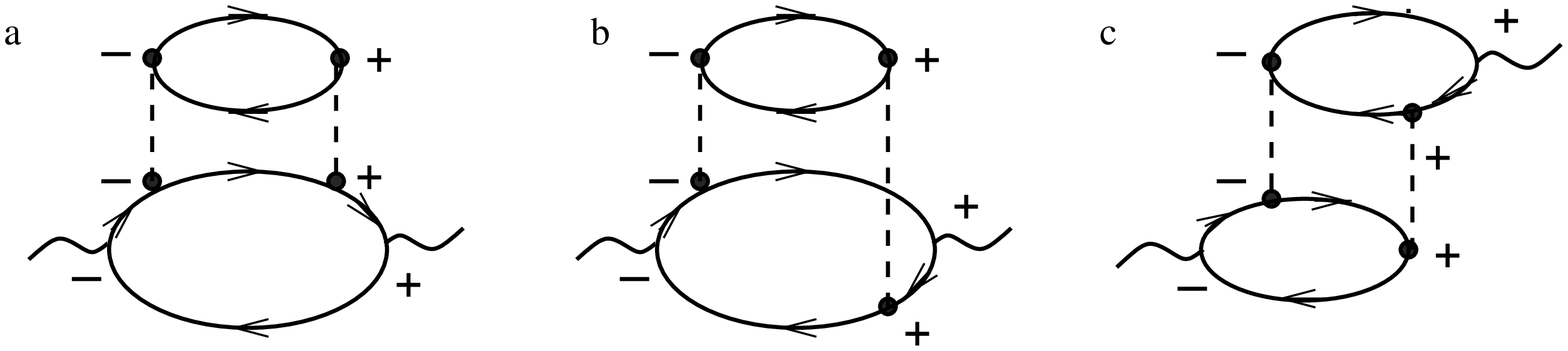}
\end{center}
{\footnotesize{Fig. 2: The Feynman diagrams for neutrino-nucleon
interaction in the 2p-2h approximation.
The vertical dashed lines correspond
to the baryon-baryon interaction and the
wavy lines to the $Z^0$ vector bosons.
Exchange diagrams are shown below in Fig. 3.
}}
\label{fig2}

\no
The analytical expression, corresponding to the Fig. 2a, is
\bea
i\Pi^{-+\, , \, a}_{\mu \nu}(q) &=&
\int\!\!\prod_{i=1}^4
\left[\frac{d^4p_i}{(2\pi)^4}\right]\frac{dk}{(2\pi)^4} (2\pi)^8
\delta(q+p_4-k-p_3) \delta(k+p_2-p_1)\Tr\left[V(k) G^{-+}(p_1)
V(k)G^{+-}(p_2)\right]\nonumber\\
&&\Tr\Bigl[\Gamma_{\mu} G^{--}(q+p_4)V(k) D^{--}(k) G^{-+}(p_3)
V(k) D^{++}(k)G^{++}(q+p_4)\Gamma_{\nu} G^{+-}(p_4)\Bigr],
\label{diag_a}
\eea
where $V(k)$ is the strong interaction vertex, which can be read-off
from eq. (\ref{VBB}).
The contribution of this diagram is readily recognized as a {\it
propagator dressing} in the $ph$ channel by means of a
self-energy corresponding to an excitation
of a single particle-hole collective mode.
The analytical expression,
corresponding to the Fig. 2b, is
\bea
i\Pi^{-+\, ,\, b}_{\mu \nu}(q) &=&
\int\!\!\prod_{i=1}^4
\left[\frac{d^4p_i}{(2\pi)^4}\right]\, \frac{dk}{(2\pi)^4}
(2\pi)^8\delta(q+p_4-k-p_3) \delta(k+p_2-p_1)
\Tr\left[V(k) G^{-+}(p_1)V(k)G^{+-}(p_2)\right]
\nonumber\\
&&\Tr\Bigl[\Gamma_{\mu} G^{--}(q+p_4)V(k) D^{--}(k) G^{-+}(p_3) \Gamma_{\nu}
V(k) D^{++}(k)G^{++}(p_3-q)G^{+-}(p_4)\Bigr].
\label{diag_b}
\eea
\noindent
The contribution of this diagram corresponds to a {\it
vertex correction} in the $ph$ channel by an
effective  interaction, which incorporates an intermediate
particle-hole collective mode excitation.
The contribution of the Fig. 2c reads
\bea
i\Pi^{-+\, , \, c}_{\mu \nu}(q) &=&
\int\!\!\prod_{i=1}^4\left[\frac{d^4p_i}{(2\pi)^4}\right]\,
\frac{dk}{(2\pi)^4}\,  (2\pi)^8\delta(q+p_4-k-p_3)\delta(k+p_2-p_1)
\nonumber\\
&&\Tr\Bigl[\Gamma_{\mu} G^{--}(q+p_4)V(k)
D^{--}(k)G^{-+}(p_3)V(k-q) G^{+-}(p_4)\Bigr]\nonumber\\
&&\Tr\left[V(k) G^{-+}(p_1)\Gamma_{\nu}G^{++}(p_1-q)
V(k-q) D^{++}(k-q) G^{+-}(p_2)\right].
\label{diag_c}
\eea
The latter diagram may be interpreted as a particle-hole
fluctuation. The diagrams $a$-$c$ are evaluated in the Appendix B.
There we show that (i) the  vector
current contributions from diagrams $a$ and $b$ mutually
cancel; (ii) the diagram $c$ does not contribute
because the axial-vector contribution involves traces over odd number
of $\sigma$-matrices and the vector-current contribution
is cancelled by an equal and of opposite sign contribution from
the diagram generated from $c$ by flipping one of the loops
upside-down; (iii) all contributions due to the Fermi-liquid
interaction cancel after summing the diagrams $a$ and $b$.
For the contraction of the trace of the neutrino current with
the polarization function we find ($i,j,=1\dots 3$)
\bea\label{CONTR}
{\cal C}_{\rm dir}(q, \bq_1, \bq_2)&=&
i\Tr(\Lambda_{ij})
\left[\Pi^{-+\, ,\, a}_{i j}(q)+
\Pi^{-+\, ,\, b}_{i j}(q)\right]  \nonumber\\
&=&{16} g_A^2G^2
\left(\frac{f_{\pi}}{m_{\pi}}\right)^4\int\!\!\prod_{i=1}^4
\left[\frac{d^4p_i}{(2\pi)^4}\right]\, \frac{d^4k}{(2\pi)^4}
G^{--}(\omega)^2 D^{--}(k)^2   \nonumber\\
&&\bk^4\left[\omega_1\omega_2 -
\frac{(\bq_1\cdot \bk)(\bq_2\cdot \bk)}{\vert k\vert^2}\right]
G^{-+}(p_1) G^{+-}(p_2)G^{-+}(p_3) G^{+-}(p_4) \nonumber\\
&&(2\pi)^4\delta(q+p_4-k-p_3)(2\pi)^4 \delta(k+p_2-p_1).
\eea
This result is valid  in the soft-neutrino and
non-relativistic baryon limits.
The second term on the r.h.s. in the square
bracket can be dropped, as it does not
contribute after the phase space   integrations. Note that the
total number of diagrams of the type $a$-$c$ is four, if one
allows for all possible relabelling of incoming and outgoing
(identical) baryons; this forfactor is equal to the symmetry factor
by which the total rate must be reduced. We do not include
these factors explicitly.

\subsection{Exchange contribution to the polarization function}

The {\it exchange} diagrams are generated from the direct ones
by means of interchanging the outgoing propagators
in a strong vertex.  There is a complete set of diagrams analogous
to $a$ and $b$  with exchanged labelling of the
hole propagators. These contribute to the contraction
\bea\label{CONTREX}
{\cal C_{\rm ex}}(q) &=& 16 g_A^2
G^2\left(\frac{f_{\pi}}{m_{\pi}}\right)^4
\omega_1\omega_2 G^{--}(\omega)^2 \int dk\,
\bk^4 \, \, D^{--}(\bk)^2\nonumber\\
&&\int\!\!\prod_{i=1}^4
\left[\frac{d^4p_i}{(2\pi)^4}\right]\,
 G^{-+}(p_1) G^{+-}(p_2)G^{-+}(p_3) G^{+-}(p_4)  \nonumber\\
&&\hspace{2cm}(2\pi)^4\delta(q+k+p_2-p_3)\delta(k+p_4-p_1),
\eea
in the soft neutrino approximation.
The skeleton diagrams which correspond to the
interference between the direct and exchange contributions
are shown  Fig. 3a-d. There are eight diagrams of each type
if one allows for all possible relabelling of the propagators.

\begin{center}
\includegraphics[height=2in,width=3.5in,angle=0]{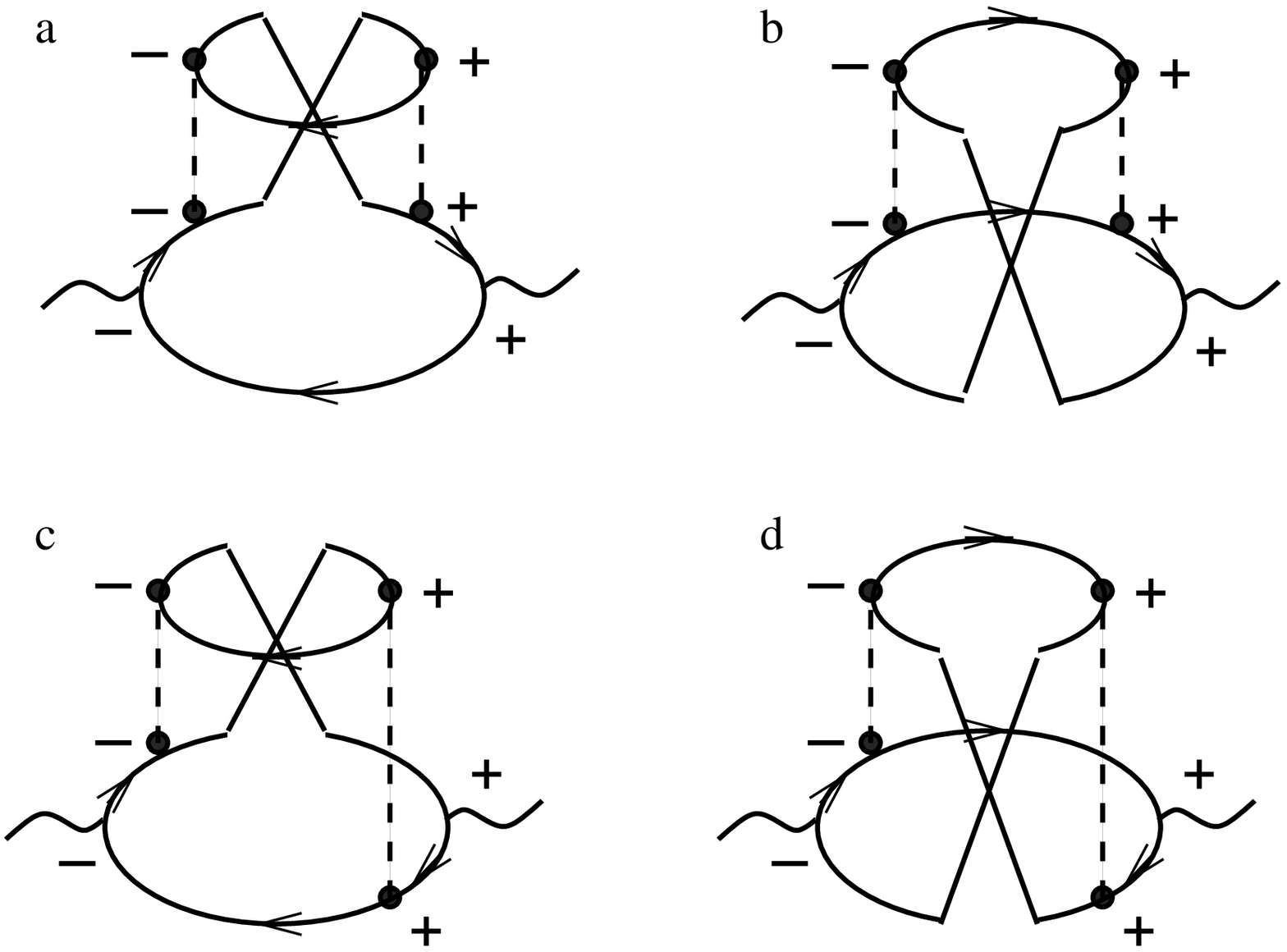}
\end{center}
{\footnotesize{Fig. 3: The exchange Feynman diagrams for baryon-baryon
interaction in the 2p-2h approximation. Conventions are the same
as in Fig. 2}}
\label{fig3}

\no
The analytical expressions for, e.g.,  the diagrams $a$ and $c$ are
\bea
i\Pi^{-+\, , \, a, {\rm ex}}_{\mu \nu}(q) &=&
\int\!\!\prod_{i=1}^4
\left[\frac{d^4p_i}{(2\pi)^4}\right]\frac{dk}{(2\pi)^4} dk' (2\pi)^8
\delta(q+p_4-k-p_3) \delta(k'+p_2-p_3) \delta(k+p_2-p_1) \nonumber\\
&& \Tr\Bigl[\Gamma_{\mu} G^{--}(q+p_4) V(k)D^{--}(k)G^{-+}(p_3)\nonumber\\
&&\hspace{2cm}
V(k') D^{++}(k) G^{+-}(p_2)V(k') G^{-+}(p_1) V(k')G^{++}(q+p_4)
\Gamma_{\nu} G^{+-}(p_4)\Bigr],
\label{diag_aex}
\eea
\bea
i\Pi^{-+\, , \, c, {\rm ex}}_{\mu \nu}(q) &=&
\int\!\!\prod_{i=1}^4
\left[\frac{d^4p_i}{(2\pi)^4}\right]\frac{dk}{(2\pi)^4} dk' (2\pi)^8
\delta(q+p_4-k-p_3) \delta(k'+p_2-p_3) \delta(k+p_2-p_1)\nonumber\\
&&\Tr\Bigl[\Gamma_{\mu} G^{--}(q+p_4) V(k)D^{--}(k)G^{-+}(p_3)  \nonumber\\
&&\hspace{2cm}
 V(k') D^{++}(k) G^{+-}(p_2)V(k') G^{-+}(p_1)\Gamma_{\nu}G^{++}(q+p_4)
V(k') G^{+-}(p_4)\Bigr],
\label{diag_cex}
\eea
and their computation is a complete analogue of that for the
direct diagrams (Appendix B).
The vector current contribution again cancels among the diagrams
$a$ and $c$ and, similarly, $b$ and $d$.
The  contribution from the interference  between the direct
and exchange diagrams to the
contraction of neutrino and baryon currents is
\bea\label{CONTREX2}
{\cal C_{\rm int}}(q) &=& 16 g_A^2
G^2\left(\frac{f_{\pi}}{m_{\pi}}\right)^4
\omega_1\omega_2 G^{--}(\omega)^2 \int dk\int dk'\,  \bk ^2\,
 \bk'^2 \, D^{--}(\bk)\, D^{--}(\bk')\nonumber\\
&&\int\!\!\prod_{i=1}^4
\left[\frac{d^4p_i}{(2\pi)^4}\right]\,
 G^{-+}(p_1) G^{+-}(p_2)G^{-+}(p_3) G^{+-}(p_4)  \nonumber\\
&&\hspace{2cm}(2\pi)^4\delta(q+p_4-k-p_3)\delta(k+p_2-p_1)
\delta(k'+p_4-p_1),
\eea
where we dropped the terms which vanish in the phase space
integrations.
The phase space integrations in the  exchange contribution
is  complicated, since the momentum
integrations do not decouple into two separate loops.
The disentanglement can be achieved by constraining
the momentum transfer in one of the pion propagators at the
value $\vert k'\vert = 2p_F$, as the main contribution to the
integral originates near this value of the momentum transfer.

\section{Quasiparticle width}

The purpose of this section is to specify
the width of baryon propagators.
To this end we carry out a full resummation in the
particle-particle ($pp$) channel by solving the scattering
$T$-matrix at finite temperatures. Our approach is based on
the Brueckner theory with the continuous energy-momentum
spectrum of  baryons. The non-perturbative
treatment of the $pp$ channel is mandatory for including
the effects of the short-range correlations due to the
repulsive part of the nucleon-nucleon force. These
correlations are then responsible for the width of quasiparticle
propagators,
$\gamma$, in our perturbation expansion in the
particle-hole ($ph$) channel.
The $ph$ interactions are dominated by the weaker long-range
part of the nucleon-nucleon interaction, which makes possible
the perturbative treatment of this channel by a truncation at
two loops.
The contour ordered $T$-matrix in the configuration space is:
\bea
\label{TMATX}
&&{\underline T}(x_1,x_2;x_3,x_4)= {\underline V}_{[pp]}(x_1,x_2;x_3,x_4)
\nonumber\\
&&\hspace{1cm}+ i {\underline V}_{[pp]}(x_1,x_2;x_3,x_4)
{\underline G}(x_7,x_5)~ {\underline G}(x_8,x_6)
{\underline T}(x_5,x_6;x_3,x_4),
\eea
where  ${\underline V}_{[pp]}(x_1,x_2;x_3,x_4)
= \sigma_z\, {V}_{[pp]}(x_1,x_2;x_3,x_4)$, is the
time-local baryon-baryon interaction in the particle-particle
channel. Note that the time locality
implies that the $pp$ propagator product
$\underline G\,\underline G \equiv
{\underline G}_{[pp]}$ should
be considered as a single matrix. The components of the scattering
amplitudes, needed for complete specification of the self-energies,
can be chosen as the retarded/advanced ones; the remaining
components are provided by the optical theorem. In the
quasiclassical limit the retarded/advanced $T$-matrices
obey the integral equation
\bea\label{TMIX}
T^{R/A}(\bp, \bp'; P) &=&
V_{[pp]}(\bp, \bp')+i\int\frac{d^3p''}{(2\pi)^3}  V_{[pp]}(\bp , \bp'')
G_{[pp]}^{R/A}(\bp'', P)T^{R/A}(\bp'', \bp', P) ,
\eea
where we kept the leading order terms in the gradient expansion
of the product ${G_{[pp]}}^{R/A} \, T^{R/A}$.
Here the subscript $[pp]$ indicates the particle-particle
channel and $\bp$, $P$ are the relative momentum and total
four-momentum respectively. The two-particle Green's function,
appearing in the kernel of equation (\ref{TMIX}), is defined as
\bea \label{G2}
G_{[pp]}^{R/A}(\bp_1, P_1) &=& \int\frac{d\omega_1}{2\pi}
\int\frac{d^4P_2}{(2\pi)^4}
\Big\{G^{>}\left(P_2/2+p_1\right) \,
G^{>}\left(P_2/2-p_1\right) \nonumber \\
&&\hspace{1cm}
-G^{<}\left(P_2/2+p_1\right)\, G^{<}\left(P_2/2-p_1\right)\Big\}
\frac{(2\pi)^3\, \delta^3({\bP_1}-{\bP_2})}{E_1-E_2\pm i\delta},
\eea
where we dropped the irrelevant dependence of the quasiclassical
functions on their center-of-mass space-time coordinates.
If the particle-hole symmetry is kept in the kernel
of the integral equation, the $T$-matrix diverges at the
critical temperature of the superfluid phase transition.
To be able to apply our computation to the low-temperature
regime (and thereby avoid the pairing instability in the
$T$-matrix) we drop the hole-hole propagators. This is a
common approximation in the Brueckner theory and is justified
in terms of the Bethe-Goldstone hole-line expansion. We treat the
intermediate state two-particle propagation in the quasiparticle
limit. Using the angle averaging procedure for the
$pp$ propagator and after partial wave expansion,
the  thermodynamic retarded $T$-matrix is given by
\bea\label{TMAT}
T^{R\alpha}_{ll'}( p,  p', P, \omega)  &=&
V^{\alpha}_{[pp]\, ll'}(p, p') \nonumber\\
&+&\frac{2}{\pi}\sum_{l''}
\int\!\! dp''\, p''^2 \, V^{\alpha}_{[pp]\, ll''}(p, p'')
\langle G_{[pp]}^R(p'', P, \omega)\rangle
T^{R\alpha}_{l''l'}(p'', p', P,\omega),
\eea
where $\alpha$ collectively denotes the quantum
numbers $(S,J,M)$ in a particular partial wave, $p$ and $P$
are the magnitudes of the relative and total momentum
respectively,  $V(p,p')$ is the bare nuclear interaction. Here
$\langle G_{[pp]}^R\rangle$ is the angle averaged two-particle propagator
\bea
\langle G_{[pp]}^R(p, P, \omega)\rangle = \int\!\frac{d\Omega}{4\pi}
\frac{\left[1-f_N(\ep(\bP/2+\bp))\right]
\left[1-f_N(\ep(\bP/2-\bp))\right]}
{\omega -\ep(\bP/2+\bp)-\ep(\bP/2-\bp)+i\delta },
\eea
with $\ep(\bp) = \epsilon_p + {\Real}\Sigma(\ep_p,\bp)$, i.e.
the intermediate state propagation is treated in the
quasiparticle approximation. The
 retarded  self-energy is given by
\bea\label{SELF_E}
\Sigma^R(p, \omega) = \frac{1}{\pi}\sum_{l\alpha}(2J+1)
\int\! dp'\, p'^2 T^{R\alpha}_{ll}(p,p';p,p';\omega+ \ep(p'))
f_N(\ep(p')),
\eea
which also defines its real and  imaginary parts.
The coupled equations (\ref{TMAT}) and (\ref{SELF_E})
are subject to normalization to the total density at a given
temperature.

\section{Phase space integrations}

Let us turn to the task of evaluating the phase space  integrals in the
expressions for the current contractions.
We substitute the Kadanoff-Baym ansatz in eq. (\ref{CONTR})
and use the identity $f_N(\ep_{1})f_N(-\ep_{2})= g(\ep_1-\ep_2)
\left[f_N(\ep_2)-f_N(\ep_1)\right]$,
which is exact in the equilibrium limit. We then find that
the contributions from each loop decouple, i.e.,
\bea
{\cal C}_{\rm dir}(q) &=& 16 g_A^2
G^2\left(\frac{f_{\pi}}{m_{\pi}}\right)^4
\omega_1\omega_2 G^{--}(\omega)^2  \int\!\! \frac{d^4k}{(2\pi)^4}
\bk ^4 D^{--}(\bk)^2\, g(\omega_k)\,
g(\omega-\omega_k)\, L(k)\, L(q-k),
\eea
where  $\omega_k=k_0$ and  the elementary loop is defined as
\bea\label{L01}
L(k) &=&\int\!\!
\frac{d^4p_1}{(2\pi)^4}\,\frac{d^4p_2}{(2\pi)^4}\,
a(p_1)a(p_2)[f_N(\ep_2)-f_N(\ep_1)] (2\pi)^4\delta(k+p_2-p_1).
\eea
The exchange contribution ${\cal C}_{\rm ex}$ leads to additional
factor of two. The interference contribution decouples only under certain
constrains.
The single loop, eq. (\ref{L01}),
can be evaluated to arbitrary order in the spectral
width in general\cite{SD}. We shall restrict to the
small quasiparticle damping limit and use the
the expansion  with respect to the width
of the spectral function given by  eq. (\ref{exp1}).

\subsubsection{Leading order}

The lowest order approximation corresponds to the quasiparticle
(i.e. zero-width) limit. The contribution from a single loop
vanishes in the time-like region of the phase space where
$\omega_k\ge\vert\bk\vert$. This result is found only if
the relativistic kinematics is applied; non-relativistic
kinematics leads to spurious terms $\propto m/q $. In the
space-like region of the phase space   the result is finite.
We carry out the energy integrations keeping only the leading
order term. Removing one of the trivial momentum delta functions
we find
\bea
L_0(k) &=&\int\!\!
\frac{d^3p}{(2\pi)^3}\,[f_N(\ep_{p})-f_N(\ep_{p+k})]
(2\pi)\delta(\omega_k+\ep_{p}-\ep_{p+k}).
\eea
The integrations can be carried out exactly 
\bea\label{LQPA}
L_0(k)&=&\int\!\!
\frac{d^3p}{(2\pi)^3}\,[f_N(\ep_{p})-f_N(\ep_{p+k})]
(2\pi)\delta(\omega_k+\ep_{p}-\ep_{p+k})
=\frac{m^{*\, 2}}{2\pi\beta\vert k\vert}
{\cal L}(\omega_k, \bk),
\eea
where $m^*$ is the effective
mass of a quasiparticle and
\bea\label{L0}
{\cal L}(\omega_k,\bk) ={\rm ln}
\Bigg\vert\frac{1+{\rm exp}\left[-\beta\left(\ep_-(k)-\mu\right)\right]}
{1+{\rm exp}\left[-\beta\left(\ep_+(k)-\mu\right)\right]}\Bigg\vert ,
\eea
with $\varepsilon_{\pm}(k)
= (\omega_k^2+\varepsilon_k^2)/4\varepsilon_k\pm\omega_k/2$. Note that
the quasiparticle loop (\ref{LQPA}) is zero in the time like
region ($\omega_k\ge\vert\bk\vert$), which sets a natural cut-off
in the phase space  integrations below.

\subsubsection{Next-to-leading order}
The next-to-leading order contribution (which is linear in $\gamma$)
is
\bea
L_1(\omega_k, k) &=& 2\int\!\!\frac{d^4 p}{(2\pi)^4}
(2\pi)\delta(\ep+\omega_k-\ep_{p+k})
\int \frac{d\omega'}{2\pi}
\gamma(p')\frac{{\cal P}}{(\omega'-\ep_{p})^2} \nn \\
&\times&
\left\{\delta(\ep -\ep_p)-\delta(\ep-\omega')
\right\}
\left[f_N(\ep) - f_N(\ep+\omega_k)\right],
\eea
where we summed the two term arising from the product of the
leading and next-to-leading order contribution to
$G^<(p)$. The angular integral can be carried out analytically
to the accuracy  ${\cal O}(\gamma^2)$. One finds
\bea\label{L1}
L_1(\omega_k, k)=-\frac{4m^{*\, 2}}{k}
\int\frac{d\ep_p}{(2\pi)^2}
\left[f_N(\ep_p)-f_N(\ep_p+\omega)\right]
\left\{{\cal Z}(\ep_p, \bk) - {\cal F} (\ep_p, \bk,\omega_k)\right\},
\eea
where the first term in the curly brackets is due to the wave-function
renormalization
\bea
 {\cal Z} (\ep_p,\bk)  =\theta(\ep_p-\ep_{\rm min})
 \int d\omega \gamma(\omega)\frac{{\cal P}}{(\omega-\ep_{p})^2},
\quad \ep_{\rm min} = \frac{(\omega_k-\ep_q)^2}{4\ep_q}.
   \eea
The second terms is the off-pole contribution and is given by
\bea
   {\cal F} (\omega_k, \bk, \ep_p)
   ={\rm arctan}\left[
   \frac{\epsilon_k-\omega_k-\mu
   +2\sqrt{\epsilon_p\epsilon_k}}{\gamma(\ep_p+\omega_k)/2}\right]
  - {\rm arctan}\left[
   \frac{\epsilon_k-\omega_k-\mu-2\sqrt{\epsilon_p\epsilon_k}}
   {\gamma(\ep_p+\omega_k)/2}
   \right].
\eea

The current contraction, which so far includes contributions
to all orders in $\gamma$, now can be decomposed in the
leading and  next-to-leading order terms with respect to
$\gamma$, employing the corresponding decomposition for the
loops. E.g. for the direct contribution one finds
\bea
{\cal C}_{\rm dir} (q) &=& 16 g_A^2
G^2\left(\frac{f_{\pi}}{m_{\pi}}\right)^4
\omega_1\omega_2 G^{--}(\omega)^2
\int \frac{d^4k}{(2\pi)^4}
\bk^4 D^{--}(\bk)^2\nn\\
&& g(\omega_k)\, g(\omega-\omega_k)\, \left[
L_0(k)\, L_0 (q-k)+L_1(k)\, L_0 (q-k) +
L_0(k)\, L_1 (q-k)\right].
\eea
The exchange and interference terms can be decomposed in
a similar manner.

\subsubsection{Neutrino emissivity}

After the preparatory work above, the computation of the neutrino
emissivity is straightforward.
We first relate the current contraction to
our original expression for the neutrino emissivity by
using the relation
$-2 g_B(q_0) {\Img} \Pi^R_{\mu\nu}(q) = i \Pi_{\mu\nu}^< (q)$.
Expression  (\ref{EMISSIVITY}) takes the form:
\bea
\epsilon_{\nu\anu}&=&
\sum_f\int\!\frac{d^3q_2}{(2\pi)^32 \omnu( q_2)}
\int\!\frac{d^3 q_1}{(2\pi)^3 2\omnu( q_1)}
\int\!\frac{d^4 q}{(2\pi)^4}
(2\pi)^4 \delta^4(q_1 + q_2 - q)
\left[\omnu(\bq_1)+\omnu(\bq_2)\right]\, {\cal C}(q),
\eea
where ${\cal C}(q)$ is the sum of the direct, exchange and
interference contributions. Let us first compute the contribution
from the direct term by substituting  eq. (\ref{CONTR})  for the
current contraction.
We carry out the integrations over the neutrino phase space
and the summation over the three neutrino flavors to find:
\bea\label{RESULT1}
\epsilon_{\nu\anu} &=&  \frac{16}{5(2\pi)^7}
g_A^2 {G_F}^2\left(\frac{f_{\pi}}{m_{\pi}}\right)^4
\int_0^{\infty}\,
d\omega\, \omega^6 G^{--}(\omega)^2
\int dk
k ^6 D^{--}(\bk)^2\nn\\
&& \int d\omega_k g(\omega_k)\, g(\omega-\omega_k)\, \left[
L_0(k)\, L_0 (q-k)+L_1(k)\, L_0 (q-k) +
L_0(k)\, L_1 (q-k)\right],
\eea
where we used $d^4k = 4\pi  k^2 dk d\omega_k$.
Normalizing the energy scales by the temperature
and the momenta by $2p_F$ we obtain
\bea\label{RESULT2}
\epsilon_{\nu\anu} &=&  \frac{32}{5(2\pi)^9} G_F^2 g_A^2
\left( \frac{f_{\pi}}{m_{\pi}} \right)^4\,
\left(\frac{m^*}{m}\right)^4\,  p_F\, I\, T^8
= 5.5 \times 10^{19}\, I_3 \,  T_9^8 ~({\rm erg~cm^{-3}~s^{-1}})
\eea
where $T_9$ is the temperature in units of $10^9$ K, $I_3$ is the
integral $I$ in units $10^3$ defined as\footnote{It is understood that the functions
of new variables are relabelled.}
\bea\label{INT}
I &=&
\int_0^{\infty}d y\, y^6 G^{--}(y)^2  {\cal Q}(y)
\int^{\infty}_0 dx  x^4 D^{--}(x)^2
\int_{-\infty}^{\infty} dz ~g(z)~g(y-z)
\Biggl\{{\cal L}(z, x) {\cal L}(y-z, x) \nn\\
&+&\frac{2}{\pi} z {\cal L}(y-z, x)\,
[{\cal F}(z, x)-{\cal Z}(z,x)]
+\frac{2}{\pi} (y-z) {\cal L}(z, x) \,
[{\cal F}(y-z, x)-{\cal Z}(y-z,x)]\Biggr\}.
\eea
The explicit dependence of eq. (\ref{RESULT2}) on the temperature
and density is the generic one \cite{CHIU_SALPETER,BAHCALL_WOLF,FLOWERS_ETAL,FRIMAN_MAXWELL,VOSKRESENSKY,PETHICK}. Additional
dependence on these parameters is contained in the integral (\ref{INT}).
Note that to avoid spurious contributions from the quasiparticle
part, (the first term in curly brackets in (\ref{INT})),
one should restrict the $z$ integration to the space-like region.
For the numerical evaluation of the neutrino emissivity we use,
following ref. \cite{FRIMAN_MAXWELL}, the free space pion propagator:
$$
D^{--}(k) = [\bk^2+m_{\pi}^2]^{-1}.
$$
The free-space approximation should be valid in the vicinity
of the nuclear saturation density. The softening
of the one-pion exchange (a precursor of the pion-condensation)
increases the neutrino emissivity by large factors
\cite{VOSKRESENSKY}. We do not attempt to accommodate this
effect as our main interest here is the role of the finite width
of quasiparticles.
The  Pauli blocking factor
\bea
{\cal Q}(y) = 30 \int_0^1 d w~
 w^2~(1-w)^2 [1-f_{\nu}(w y)] [1-f_{\anu}((1-w)y)],
\eea
accounts for the occupation of neutrino and anti-neutrino final states.
In the dilute (anti-)neutrino limit $\beta\mu_{\nu_f}\ll 1$
(where $\mu_{\nu_f}$ is the  chemical potential of neutrinos of
flavor $f$) ${\cal Q}(y) =1$.

In the low-temperature  limit
 ${\cal L}(z) = z$ and the $z$-integration
decouples from the $x$-integration. On imposing $\gamma(\omega)\to 0$
(quasiparticle limit)
one finds that   ${\cal F}=0$  and $G^{--}(\omega)=\omega^{-2}$.
Then the $z$ integration can be carried analytically upon
dropping the wave-function renormalization contribution:
\be
\int_{-\infty}^{\infty}\!\! dz ~g(z)\, g(y-z)~ z ~(y-z)
 = \frac{y~(y^2+4\pi^2)}{6~(e^y-1)} .
\ee
After these manipulations eq. (\ref{RESULT2})  reduces
to Friman and Maxwell's result [ref. \cite{FRIMAN_MAXWELL}, eq. (47)].
The numerical coefficient in eq. (\ref{RESULT2}), however,
is by a factor 3 larger, since Friman and Maxwell  do not carry out
the summation over the three neutrino flavors at that stage.

The contribution from the exchange current contraction, eq.
(\ref{CONTREX}), leads to a factor of 2 in the integral (\ref{INT}).
The contribution of the interference term, in the approximation where
one of the momentum transfers is fixed at the characteristic value
$2p_F$, is
\bea\label{INT2}
I_{\rm int} &=&
\int_0^{\infty}d y\, y^6 G^{--}(y)^2 {\cal Q}(y)
\int^{\infty}_0 dx  x^2 D^{--}(x)  D^{--}(1)
\int_{-\infty}^{\infty} dz ~g(z)~g(y-z)
\Biggl\{{\cal L}(z, x) {\cal L}(y-z, 1) \nn\\
&+&\frac{2}{\pi} z {\cal L}(y-z, x)\,
[{\cal F}(z, 1)-{\cal Z}(z,1)]
+\frac{2}{\pi} (y-z) {\cal L}(z, 1) \,
[{\cal F}(y-z, x)-{\cal Z}(y-z,x)]\Biggr\}.
\eea

\section{Results}
The numerical calculations
were carried out for  pure neutron matter using the
Paris $NN$ interaction keeping $J\le 4 $ partial waves.
Fig. 4 displays the real part of
the on-shell self-energy and the
half width of the spectral function as a function of the particle
momentum for several values of the temperature at the
saturation density $n_s=0.17$ fm$^{-3}$.
The  width of the quasiparticle propagators can be parametrized
in terms of the reciprocal of the quasiparticle life time in the
Fermi-liquid theory (damping of the zero sound):
\be\label{ZERO_SOUND}
\gamma = a T^2\left[1+\left(\frac{\omega}{2\pi T} \right)^2 \right],
\ee
where $a$ is a density dependent phenomenological parameter. The parabolic
dependence of the width on the frequency is justified for temperatures
below 30 MeV in the range of the densities
$n_s \le n\le 2 n_s$. The quadratic dependence
of $\gamma$ on the temperature breaks down at slightly lower temperatures.
The value of the parameter $a$ weakly depends on the density and is
approximately 0.2 MeV$^{-1}$.

\begin{center}
\includegraphics[height=6.in,width=5.in,angle=0]{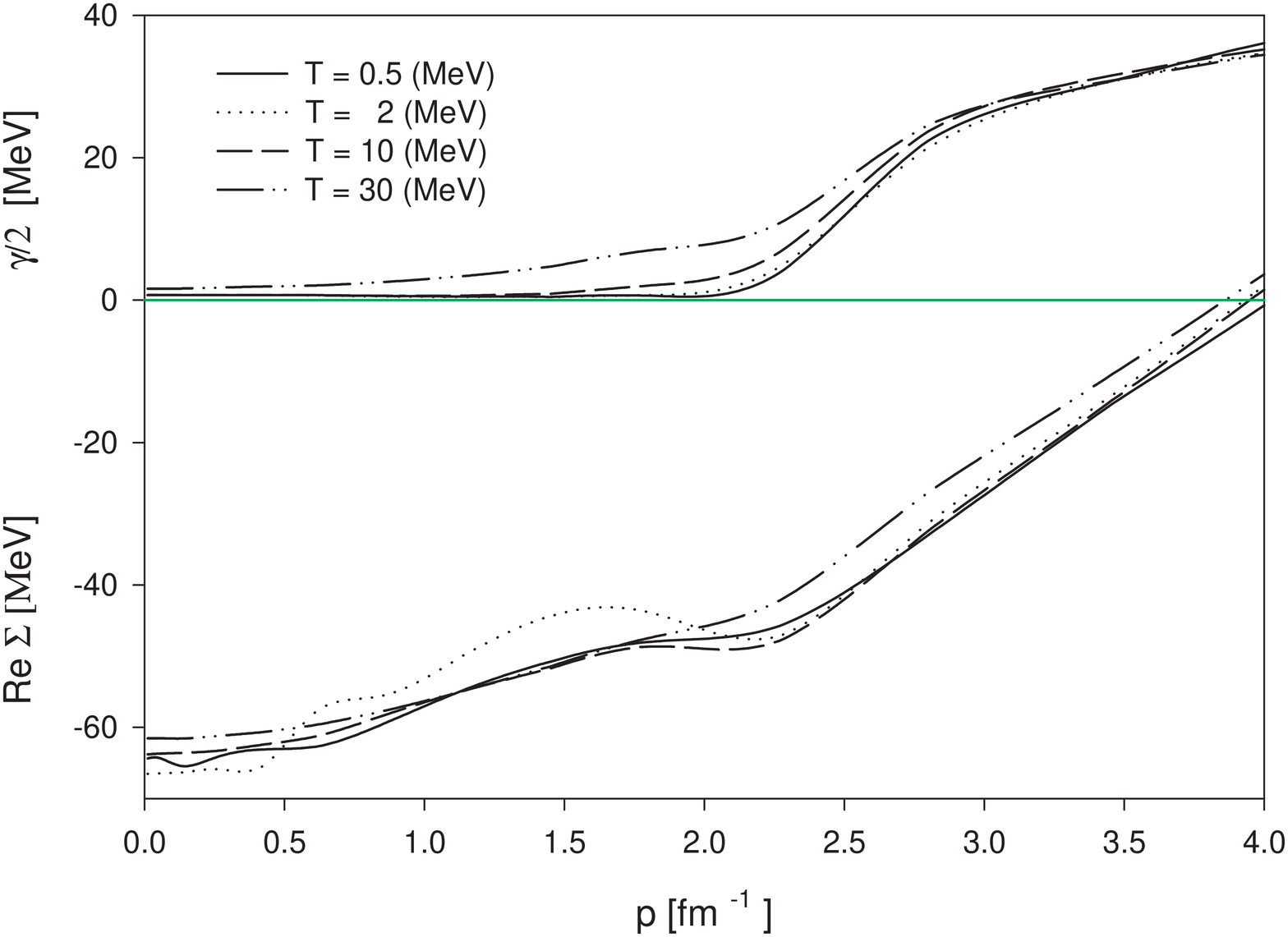}
\end{center}
{\footnotesize{Fig. 4: The real part of the on-shell
self-energy and the half-width
as a function of particle momentum at the saturation density $n_s=0.17$
fm$^{-3}$ for different temperatures; the zero temperature
Fermi momentum is 1.7 fm $^{-1}$.}}
\label{fig4}

\no
The emergent neutrino spectrum can be caracterized by their spectral
function  
\bea\label{SPEC}
S(y) &=&
 G^{--}(y)^2  {\cal Q}(y)
\int^{1}_0 dx  x^4 D^{--}(x)^2
\int_{-\infty}^{\infty} dz ~g(z)~g(y-z)
\Biggl\{{\cal L}(z, x) {\cal L}(y-z, x) \nn\\
&+&\frac{2}{\pi} z {\cal L}(y-z, x)\,
[{\cal F}(z, x)-{\cal Z}(z,x)]
+\frac{2}{\pi} (y-z) {\cal L}(z, x) \,
[{\cal F}(y-z, x)-{\cal Z}(y-z,x)]\Biggr\}, 
\eea
which is depicted in Fig. 5. 
The values of the integral are shown as
a function of neutrino frequency at $T = 20$ MeV
and the saturation density $n_s=0.17$
fm$^{-3}$ in the limit  of vanishing width ({\it dashed line}),
including the leading order contribution in the width 
({\it dashed-dotted line}) and full non-perturbative result
({\it solid line}). The energy carried by neutrinos is 
of order of $\omega \sim 5T$ in all three cases, 
as the peak in the spectral function is independent 
of the approximation to the width of the propagators.
The integral $I_3$ is show in Fig. 6.
The finite width of propagators 
leads to a suppression of the bremsstrahlung rate 
as a result of the LPM effect. 
Keeping the full non-perturbative
expression for the causal propagators enhances the value of the
integral, as the higher order terms contribute additively to the
leading order result.
The LPM effect sets in roughly when
$\omega \sim \gamma$. As neutrinos are produced thermally,
the onset temperature of the LPM effect is
of the order of $\gamma$. Equation (\ref{ZERO_SOUND})
shows that the value of the parameter $a$ controls  the onset temperature
which turns out of the order of 5 MeV in agreement with the
previous results of
refs. \cite{RAFFELT_SECKEL,JANKA,RAFFELT,HANESTAD}
and  our numerical computation  (see Fig. 6).

\begin{center}
\includegraphics[height=4.in,width=4.in,angle=-90]{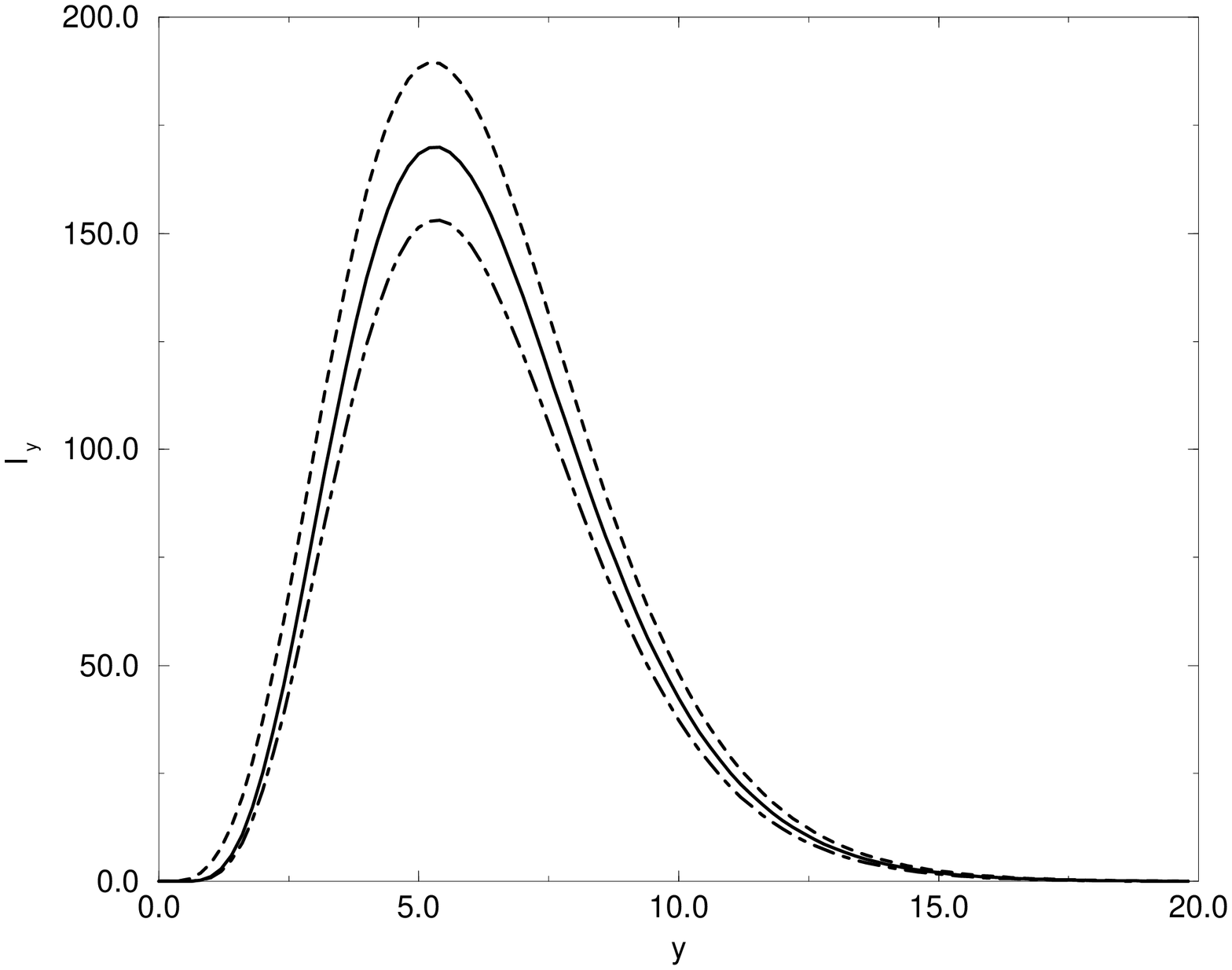}
\end{center}
{\footnotesize{Fig. 5: The neutrino spectral function 
(\ref{SPEC}) at the temperature $T=20$ MeV and 
density $n_s = 0.16$ fm$^{-3}$. 
The dashed curve is the zero width limit, the 
dashed-dotted curve includes only the leading order in $\gamma$ contribution
from the causal propagator, the solid curve is the full non-perturbative
result.
}}
\label{fig5}

\begin{center}
\includegraphics[height=4.in,width=4.in,angle=-90]{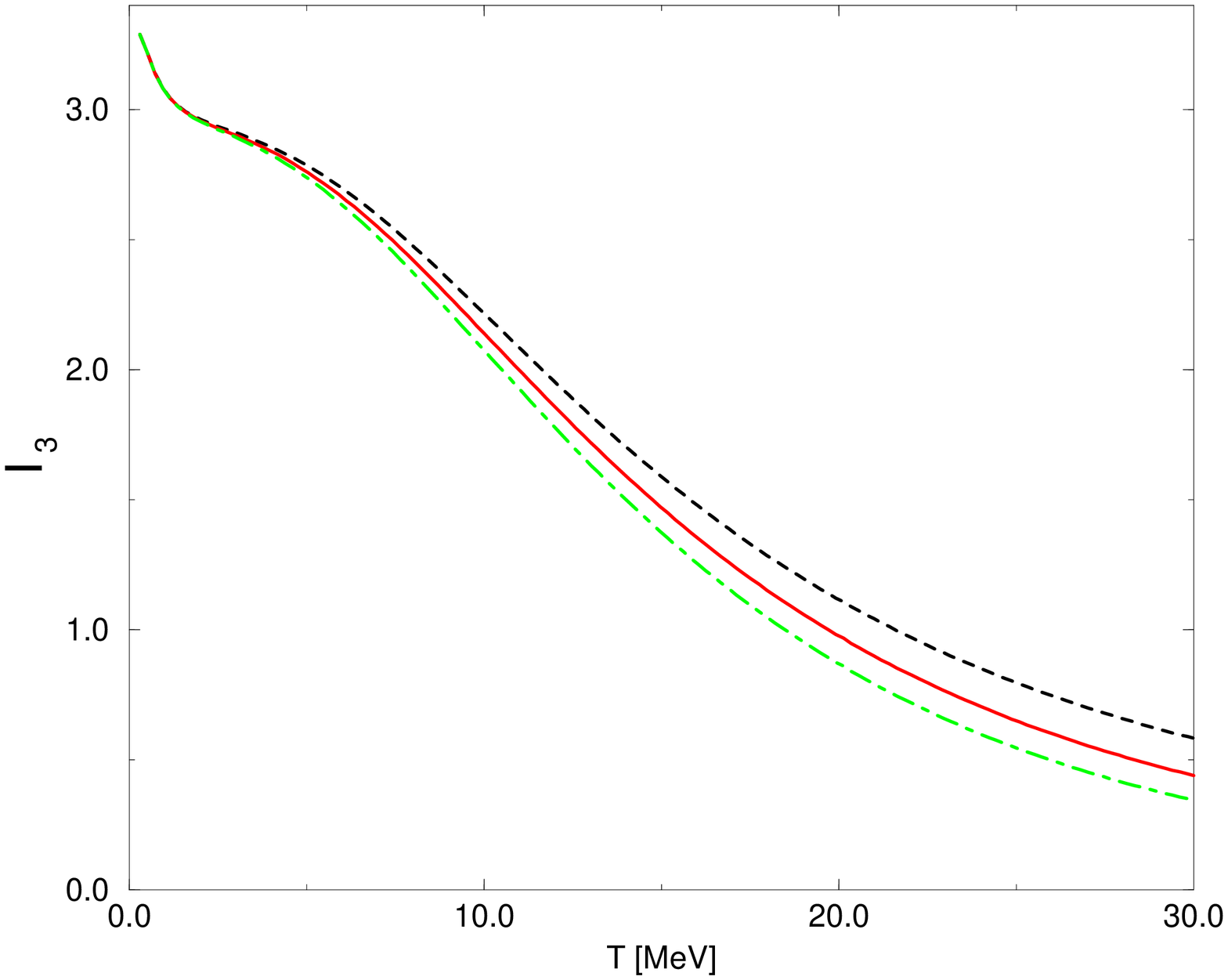}
\end{center}
{\footnotesize{Fig. 6: The integral (\ref{INT}) (including the exchange
terms) as a function of temperature
at the density $n_s = 0.16$ fm$^{-3}$.
The dashed curve is the zero width limit, the 
dashed-dotted curve includes only the leading order in $\gamma$ contribution
from the causal propagator, the solid curve is the full non-perturbative
result.
}}
\label{fig6}

\section{Conclusions}

In this work we formulated a transport theory for neutrinos
in the framework of real-time Green's functions
formalism, with particular attention to the collision integrals
for the neutrino-pair bremsstrahlung. The main focus was
a first principle  calculation of the bremsstrahlung emissivity
including the width of propagators. This allows to answer
questions, not covered by the semi-phenomenological theory,
such as the magnitude of the contribution
of higher order terms in the expansion with respect to the
quasiparticle width or the cancellation
of the  vector current contribution at all orders
in the quasiparticle width.
Even though the expression for the emissivity, which follows
from our quasiclassical transport equation, is the same as the one found
in the linear response theory, it is valid under conditions
arbitrary far from equilibrium. This is particularly important
in the regime where the  neutrinos decouple from matter and
their distribution function strongly deviates from the Fermi-Dirac
form.

The central quantity of the theory is
the particle-hole polarization tensor
in the $ph$ channel truncated at two loops.
The $pp$ channel is treated non-perturbatively
within the finite temperature Brueckner theory.
We find that the only contribution
to the bremsstrahlung rate
comes from the contraction of the tensor force with the
axial vector current to all orders in the
quasiparticle width. Other contributions, which arise
from the contraction of the  Fermi-liquid
type interaction with the axial vector current
and the contraction of the net strong interaction with the
vector current, cancel when we take the sum
of the diagrams corresponding to vertex corrections
and propagator renormalization in the $ph$ channel. Thereby the
vector current conservation is established at all orders
in the quasiparticle width. These
cancellations are independent of the approximations
to the propagators and are effective both in the quasiparticle limit
and beyond.
The three ingredients crucial to the cancellations are:
(i) the anti-commutation of the
tensor force with the axial vector current, (ii) the odd parity
of the causal propagator under the exchange of its energy
argument, (iii) the soft neutrino and non-relativistic
baryon approximations.

Our numerical evaluation of the neutrino emissivity
of hot neutron matter, carried out
at two loops, shows that the LPM-type suppression sets in
at temperatures $T\ge \gamma$, in agreement with the
previous work limited to the
first order terms in the quasiparticle width (see ref. \cite{RAFFELT}
and references therein). The higher order terms
enhance the magnitude of the neutrino emissivity compared to the
leading order result. The non-perturbative result, however, is still
suppressed as compared to the quasiparticle limit.  

Our formalism can be extended in various ways. One obvious
extension is allowing for two different chemical potentials
of scattering baryons. This will include the Urca process
(the $\beta$-decay in the second order in the virial expansion)
and the effects of the Pauli spin-paramagnetism,
which become important in strong magnetic fields.
The formalism can be 
adapted, with minor changes, for a computation
of the space-like analogous of the bremsstrahlung and,
in particular, the neutrino opacities of
the supernova matter.
The perturbative scheme, employed here,  itself requires
further improvements in several  direction, numerically
the most  important one being the renormalization of the
one-boson exchange interaction in the $ph$ channel.

\section*{Acknowledgements}

This work has been supported by the Stichting voor
Fundamenteel Onderzoek der Materie
with financial support from the Nederlandse Organisatie
voor Wetenschappelijk Onderzoek.
A.S. thanks the Institute for Nuclear Theory at the
University of Washington for its hospitality and the
Department of Energy for partial support during the
completion of this work.


\begin{appendix}
\section{Real-time Green's functions}

The six Green's functions  of the non-equilibrium theory are
not independent. For completeness we summarize here the linear
relations among them, which can be easily verified from their
definitions. The four components of the matrix Green's function
are related to each other by the relations
\bea
&&     S^{--}(x_1,x_2)= \theta(t_{1}-t_{2}) S^{+-}(x_1, x_2) +
             \theta(t_{2}-t_{1}) S^{-+}(x_1, x_2) ,  \\
\label{GF1}
&&     S^{++}(x_1,x_2)= \theta(t_{2}-t_{1}) S^{+-}(x_1,x_2) +
             \theta(t_{1}-t_{2}) S^{-+}(x_1,x_2), \\
\label{GF2}
&&     S^{--}(x_1,x_2) + S^{++}(x_1,x_2)= S^{-+}(x_1,x_2) + S^{+-}(x_1,x_2).
\label{GF3}
\eea
Following Hermitian conjugation relations hold:
\bea
S^{--}(x_1,x_2)= - S^{++*}(x_2,x_1),
\label{HC1}\\
S^{-+}(x_1,x_2)= - S^{-+*}(x_2,x_1),
\label{HC2}\\
S^{--}(x_1,x_2)= - S^{+-*}(x_2,x_1).
\label{HC3}
\eea
The retarded and advanced Green's functions are
related to the components of the matrix Green's function
via the relations
\bea
S^{R}(x_1,x_2) &=& \theta(t_{1}-t_{2})
\left[S^{+-}(x_1,x_2)-S^{-+}(x_1,x_2)\right]\nonumber \\
&=&S^{--}(x_1,x_2)-S^{-+}(x_1,x_2)=S^{+-}(x_1,x_2)-S^{++}(x_1,x_2),
\label{APP:RET}   \\
S^{A}(x_1,x_2) &=& - \theta(t_{2}-t_{1})
\left[S^{+-}(x_1,x_2)-S^{-+}( x,y ) \right] \nonumber \\
&=& S^{--}(x_1,x_2)-S^{+-}(x_1,x_2)=S^{-+}(x_1,x_2)-S^{++}(x_1,x_2).
\label{APP:ADV}
\eea
They are Hermitian conjugates, i.e.
\be
S^A(x_1,x_2) = S^{R*}(x_1,x_2).
\ee
In addition, we note that in the momentum representation
they satisfy the equations
\bea\label{HC4}
S^{--}(\omega, \bp) = - \left[ S^{++}(\omega,\bp)\right]^* ,\quad
S^{A}(\omega, \bp)  =  \left[ S^{R}(\omega,\bp)\right]^*.
\eea
 The relations above are valid for the baryon and pion propagators
 in general, and we do not repeat them here.

 Similar relations hold among the self-energies. These can be identified
 by performing  a unitary orthogonal transformation affected by the
 matrix $R = (1+i\sigma_y)/2$ by means of formula $S' = R^{-1}SR$.
 The  form of the original Dyson equation in the matrix form (\ref{DYSON1})
 is invariant against the transformation,
 \be
 \underline{S'}(x_1,x_2) = \underline{S'}_0(x_1,x_2)
       + \underline{S'}_0(x_1,x_3)
         \underline{\Omega'}(x_3,x_2) \underline{S'}(x_2,x_1)
 \ee
 where the primed quantities have the ``triangular'' form
 \be
  S'_{12} = \left( \begin{array}{cc}
                              0 & S^{A}_{12} \\
                              S^{R}_{12} & S^{K}_{12}
                           \end{array} \right),
                           \quad
 \Omega'_{12} =\left( \begin{array}{cc}
                              \Omega^K_{12} & \Omega^{R}_{12} \\
                              \Omega^{A}_{12} & 0\end{array} \right).
\ee
where
\be
  S^K(x_1,x_2) = S^c(x_1,x_2)+S^a(x_1,x_2) =  S^>(x_1,x_2)+S^<(x_1,x_2),
\ee
 \be
\Omega^R(x_1,x_2)  = \Omega^c(x_1,x_2) + \Omega^{<}(x_1,x_2),\quad
     \Omega^A(x_1,x_2) =  \Omega^c(x_1,x_2) + \Omega^{>}(x_1,x_2),
  \ee
  \be
   \Omega^K(x_1,x_2) = \Omega^c(x_1,x_2) +\Omega^a(x_1,x_2)
   =-\Omega^>(x_1,x_2)-\Omega^<(x_1,x_2).
 \ee

\section{Details of the computation of the polarization function}

As an example we compute here the direct contribution to the
polarization function, represented by the diagrams $a$ and $b$
in Fig. 2. The cancellation among the various contributions
from these diagrams does not depend on the details of the structure
of the baryon propagators (quasiparticle or dressed), but solely on the
odd parity of the causal Green's function with respect to a
change of the sign of the energy argument in the soft neutrino
approximation.

In the first step we substitute the vertices. As the contribution
of the Landau Fermi-liquid part of the interaction
will cancel out, to save space, we shall drop its contribution
from the outset. For the diagrams $a$ and $b$ (excluding the
factors for the topologically equivalent diagrams) we find
\bea
i\Pi^{-+, a}_{\mu \nu}(q) &=&
\left(\frac{G}{2\sqrt{2}}\right)^2\left(\frac{f}{m_{\pi}}\right)^4
\int\!\!\prod_{i=1}^4
\left[\frac{d^4p_i}{(2\pi)^4}\right]\, \frac{dk}{(2\pi)^4}\nonumber\\
&&\Tr\Bigl[\left(\delta_{\mu 0}
-g_A\delta_{\mu i}\sigma_i\right)
G^{--}(q+p_4)\left(\bsigma\cdot\bk\right) D^{--}(k)\nonumber \\
&&\hspace{2cm} G^{-+}(p_3)
\left(\bsigma\cdot\bk\right) D^{++}(k)
G^{++}(q+p_4)\left(\delta_{\nu 0}-g_A
\delta_{\nu j}\sigma_j\right) G^{+-}(p_4)\Bigr]\nonumber\\
&&\Tr\left[\left(\bsigma\cdot\bk\right)
G^{-+}(p_1)\left(\bsigma\cdot\bk\right)G^{+-}(p_2)\right]
(2\pi)^8\delta(q+p_4-k-p_3)\delta(k+p_2-p_1),
\label{diag_a_app}
\eea
\bea
i\Pi^{-+, b}_{\mu \nu}(q) &=&  \left(\frac{G}{2\sqrt{2}}\right)^2
\left(\frac{f}{m_{\pi}}\right)^4
\int\!\!\prod_{i=1}^4
\left[\frac{d^4p_i}{(2\pi)^4}\right]\, \frac{dk}{(2\pi)^4}\nonumber\\
&&\Tr\Bigl[\left(\delta_{\mu 0}-g_A\delta_{\mu i}\sigma_i\right)
G^{--}(q+p_4)\left(\bsigma\cdot\bk\right) D^{--}(k)\nonumber\\
&&\hspace{2cm} G^{-+}(p_3) \left(\delta_{\nu 0}-g_A\delta_{\nu j}\sigma_j\right)
\left(\bsigma\cdot\bk\right) D^{++}(k)G^{++}(p_3-q)G^{+-}(p_4)\Bigr]\nonumber\\
&&\Tr\left[\left(\bsigma\cdot\bk\right) G^{-+}(p_1)
\left(\bsigma\cdot\bk\right)G^{+-}(p_2)\right](2\pi)^8\delta(q+p_4-k-p_3)
\delta(k+p_2-p_1).
\label{diag_b_app}
\eea
Next we apply the approximation (\ref{DENOM_EXP}) to the causal and acausal
Green's functions and fix their momenta at the corresponding
Fermi momentum.  Combining the diagrams $a$ and $b$, we find
\bea
i\Pi^{-+, a}_{\mu \nu}(q)+ i\Pi^{-+, b}_{\mu \nu}(q) &=&
\left(\frac{G}{2\sqrt{2}}\right)^2\left(\frac{f}{m_{\pi}}\right)^4
 \int\!\!\prod_{i=1}^4
\left[\frac{d^4p_i}{(2\pi)^4}\right]\, \frac{dk}{(2\pi)^4}\nonumber\\
&&G^{--}(\omega)^2 D^{--}(k)^2
G^{-+}(p_1) G^{+-}(p_2)G^{-+}(p_3) G^{+-}(p_4)  \nonumber\\
&&\Tr\Bigl\{\left(\delta_{\mu 0}-g_A\delta_{\mu i}\sigma_i\right)
\left(\bsigma\cdot\bk\right)
\left(\bsigma\cdot\bk\right)
\left(\delta_{\nu 0}-g_A\delta_{\nu j}\sigma_j\right)\nonumber\\
&& -\left(\delta_{\mu 0}-g_A\delta_{\mu i}\sigma_i\right)
\left(\bsigma\cdot\bk\right)
 \left(\delta_{\nu 0}-g_A\delta_{\nu j}\sigma_j\right)
\left(\bsigma\cdot\bk\right)\Bigr\}\nonumber\\
&&\Tr\left[\left(\bsigma\cdot\bk\right)
\left(\bsigma\cdot\bk\right)\right]
(2\pi)^8\delta(q+p_4-k-p_3) \delta(k+p_2-p_1),
\label{diag_sum}
\eea
where we used the conjugation relation (\ref{HC4}).
The terms under the trace
$\propto \delta_{0\mu}, \delta_{0\nu}$
vanish. The $\Pi_{00}$ component of the polarization
is hence zero and the vector current is conserved.
The remainder simplifies to
\bea\label{APP:B1}
&&i\Pi^{-+,a}_{i j}(q)+i\Pi^{-+,b}_{i j}(q)=
g_A^2 \left(\frac{G}{2\sqrt{2}}\right)^2
\left(\frac{f}{m_{\pi}}\right)^4\int\!\!\prod_{i=1}^4
\left[\frac{d^4p_i}{(2\pi)^4}\right]\,\frac{dk}{(2\pi)^4}\nonumber\\
&&[G^{--}(\omega)^2 D^{--}(k)^2
G^{-+}(p_1) G^{+-}(p_2)G^{-+}(p_3) G^{+-}(p_4)  \nonumber\\
&&\Tr\Bigl[\sigma_{i} \left(\bsigma\cdot\bk\right)
\left(\bsigma\cdot\bk\right) \sigma_{j}
 -\sigma_{i} \left(\bsigma\cdot\bk\right)
\sigma_{j}\left(\bsigma\cdot\bk\right)\Bigr]
\Tr\left[\left(\bsigma\cdot\bk\right)
\left(\bsigma\cdot\bk\right)\right]\nonumber\\
&&
(2\pi)^8\delta(q+p_4-k-p_3) \delta(k+p_2-p_1).
\eea
The computation of the trace using the $\bsigma$-algebra gives
\bea
 \Tr\left[\left(\bsigma\cdot\bk\right)
\left(\bsigma\cdot\bk\right)\right]
\Tr\Bigl[\sigma_{i} \left(\bsigma\cdot\bk\right)
\left(\bsigma\cdot\bk\right) \sigma_{j}
 -\sigma_{i} \left(\bsigma\cdot\bk\right)
\sigma_{j}\left(\bsigma\cdot\bk\right)\Bigr]
= 8k^2 \left(k^2\delta_{ij} -k_ik_j \right).
\eea
The contraction of the polarization tensor with the trace
of neutrino currents,
given by
\be
\Tr\Lambda_{ij} = 8 \left[q_{1i} q_{2j} + q_{1j} q_{2i}
+\left(\omega_1\omega_2+\bq_1\cdot\bq_2\right)\delta_{ij}
+\epsilon_{injm} q_{1}^nq_{2}^m\right],
\ee
leads to
\be\label{APP:B2}
  8k^2 \Tr\Lambda_{ij}\left(k^2\delta_{ij} -k_ik_j \right)
  = 128 k^4 \left[\omega_1\omega_2 -
  \frac{(\bq_1\cdot \bk)(\bq_1\cdot \bk)}{k^2} \right].
\ee
Combining eqs. (\ref{APP:B1}) and (\ref{APP:B2})
we recover eq. (\ref{CONTR}).

Let us turn to the fluctuation diagram in  Fig. 1c. From the
original diagram one can generate three additional ones
by turning each of the loops upside-down.
Let us combine the diagram in Fig. 1c with its counterpart, say $c'$,
which results from $c$ by turning the  upper loop  upside-down.
The analytical expression corresponding to their sum is
\bea
i\Pi^{-+, c}_{\mu \nu}(q)+ i\Pi^{-+, c'}_{\mu \nu}(q) &=&
\left(\frac{G}{2\sqrt{2}}\right)^2\left(\frac{f}{m_{\pi}}\right)^4
\int\!\!\prod_{i=1}^4\left[\frac{d^4p_i}{(2\pi)^4}\right]\,
\frac{dk}{(2\pi)^4}\nonumber\\
&&G^{--}(\omega)^2 D^{--}(k)^2  G^{-+}(p_1)
G^{+-}(p_2)G^{-+}(p_3)G^{+-}(p_4)\nonumber\\
&&\Biggr\{\Tr\Bigl[\left(\delta_{\mu 0}
-g_A\delta_{\mu i}\sigma_i\right)
\left(\bsigma\cdot\bk\right)
\left(\bsigma\cdot\bk\right)
\Bigr]\Tr\left[\left(\bsigma\cdot\bk\right)
\left(\delta_{\nu 0}-g_A\delta_{\nu j}\sigma_j\right)
\left(\bsigma\cdot\bk\right)\right]\nonumber\\
&-&
\Tr\Bigl[ \left(\bsigma\cdot\bk\right)
\left(\delta_{\mu 0}-g_A\delta_{\mu i}\sigma_i\right)
\left(\bsigma\cdot\bk\right)\Bigr]
\Tr\left[\left(\bsigma\cdot\bk\right)
\left(\delta_{\nu 0}-g_A\delta_{\nu j}\sigma_j\right)
\left(\bsigma\cdot\bk\right)\right]
\Biggr\}\nonumber\\
&&(2\pi)^8\delta(q+p_4-k-p_3)\delta(k+p_2-p_1),
\label{diag_1a1}
\eea
where we dropped $\bq$ compared with $\bk$ in the strong interaction
vertex. The contribution due to the axial-vector
current vanishes because  the traces are over odd number of
$\bsigma$ matrices; the contribution due
the vector current cancels as these are identical for diagrams
$c$ and $c'$ and are of opposite sign.
\end{appendix}


\begin{thebibliography}{99}
\bibitem{CHIU_SALPETER} H. Y. Chiu and E. E. Salpeter,
                        Phys. Rev. Lett. {\bf 12}, 413 (1964).
\bibitem{BAHCALL_WOLF}  J. N. Bahcall and R. A. Wolf,
                        Phys. Rev. Lett. {\bf 14}, 343 (1965);
                        Phys. Rev. {\bf 140}, B1452 (1965).
\bibitem{FLOWERS_ETAL}  E. G. Flowers,  P. G. Sutherland and
                        J. R. Bond, Phys. Rev. D {\bf 12}, 315 (1975).
\bibitem{FRIMAN_MAXWELL}B. L. Friman and O. V. Maxwell,
                        Ap. J. {\bf 232}, 541 (1979).
\bibitem{VOSKRESENSKY}  D. N. Voskresensky and A. V. Senatorov,
                       Sov. Phys. JETP {\bf 63}, 885 (1986);
                       Sov. J. Nucl. Phys. {\bf 45}, 411 (1987).
\bibitem{PETHICK} C. J. Pethick, Rev. Mod. Phys. {\bf 64}, 1133 (1992).
\bibitem{RAFFELT_SECKEL} G. Raffelt and D. Seckel, Phys. Rev. Lett.
                       {\bf 67}, 2605 (1991); Phys. Rev. D {\bf 52},
                       1780 (1995).
\bibitem{JANKA} H.-T. Janka, W. Keil, G. Raffelt, D. Seckel, Phys. Rev.
                       Lett. {\bf 76}, 2621 (1996).
\bibitem{RAFFELT} G. Raffelt, {\it Stars as  Laboratories for
                           Fundamental Physics}
                           (Univ. Chicago Press, Chicago, 1996).
\bibitem{HANESTAD} S. Hannestad and G. Raffelt, Astrophys. J. {\bf 507},
                  339 (1998).
\bibitem{LP} L. D. Landau and I. Ya. Pomeranchuk, Dokl. Akad. Nauk
             SSSR {\bf 92} 535 (1953); {\it ibidem} {\bf 92}, 375 (1953).
A. B. Migdal, Phys. Rev. {\bf 103}, 1811   (1956).
\bibitem{KNOLL_VOSKRESENSKY} J. Knoll and D. N. Voskresensky,
                             Ann. Phys. (NY) {\bf 249}, 532 (1996).
\bibitem{IWAMOTO_PETHICK} N. Iwamoto and C. J. Pethick, Phys. Rev. D {\bf 25},
                          313   (1982).
\bibitem{SAWYER} R. F. Sawyer, Phys. Rev. C {\bf 40}, 865 (1989);
                      Phys. Rev. Lett., {\bf 75}, 2260 (1995).
\bibitem{HOROWITZ} C. J. Horowitz and K. Wehrberger,
                         Nucl. Phys. A {\bf 531}, 665 (1991).
\bibitem{HAENSEL} P. Haensel and A. J. Jerzak, Astron. Astrophys. {\bf 179},
                   127 (1987).
\bibitem{BURROWS} A. Burrows and R. F. Sawyer, Phys. Rev. C
                         {\bf 58}, 554 (1998).
\bibitem{REDDY_ETAL} S. Reddy, M. Prakash and J. M. Lattimer,
                     Phys. Rev. D {\bf 58}, 013009 (1998).
\bibitem{BRAATEN} E. Braaten and R. D. Pisarski, Nucl. Phys. B {\bf 337},
               569 (1990).
\bibitem{SD} A. Sedrakian and A. Dieperink, Phys. Lett B {\bf 463}, 145 (1999).
\bibitem{KADANOFF_BAYM} L. P. Kadanoff and G. Baym,
                        {\it Quantum Statistical Mechanics}
                        (Benjamin, New York,  1962).
\bibitem{MALFLIET} W. Botermans and R. Malfliet, Phys. Rep. {\bf 198},
                   115 (1990).
\end{thebibliography}
\end{document}